\documentclass[preprintnumbers, floatfix, onecolumn, 10pt, letterpaper, superscriptaddress,nofootinbib,aps,prd]{revtex4-2}
\usepackage{graphicx}
\usepackage{subcaption}
\usepackage{microtype}
\usepackage{amsmath}
\usepackage{amssymb}
\usepackage{url}
\usepackage{xcolor}
\usepackage{mathrsfs}
\usepackage{latexsym}
\usepackage{ragged2e}
\usepackage{textcomp}
\usepackage{phaistos}
\usepackage{booktabs}
\usepackage{hyperref}
\makeatletter
\renewcommand\@makefnmark{\hbox{\@textsuperscript{\normalfont\color{purple}\@thefnmark}}}
\renewcommand\@makefntext[1]{%
\parindent 1em\noindent
\hb@xt@1.8em{%
\hss\@textsuperscript{\normalfont\@thefnmark}}#1}
\makeatother

\DeclareCaptionJustification{justified}{\leftskip=0pt \rightskip=0pt \parfillskip=0pt plus 1fil}
\definecolor{vividviolet}{rgb}{0.62, 0.0, 1.0}
\definecolor{amaranth}{rgb}{0.9, 0.17, 0.31}
\definecolor{palatinateblue}{rgb}{0.15, 0.23, 0.89}
\definecolor{brightpink}{rgb}{1.0, 0.0, 0.5}
\definecolor{cornflowerblue}{rgb}{0.39, 0.58, 0.93}
\definecolor{deepcarminepink}{rgb}{0.94, 0.19, 0.22}
\definecolor{radicalred}{rgb}{1.0, 0.21, 0.37}
\colorlet{RED}{red}

\hypersetup{ linktoc=all,
colorlinks, linkcolor={palatinateblue},
citecolor={brightpink}, urlcolor={amaranth}
}

\makeatletter
\def\@fnsymbol#1{\ensuremath{\ifcase#1
\or $\textleaf$ \or $\PHplaneTree$ \or $\PHrosette$ \or $\PHvine$
\else\@ctrerr\fi}}%
\makeatother

\begin{document} 
\title{Spin precession frequencies of a test gyroscope around a naked singularity and quasi-periodic oscillations}

\author{Tehreem Zahra}
\email{tehreemzahra971@gmail.com}
\affiliation{School of Natural Sciences, National University of Sciences and Technology, H-12, 44000-Islamabad, Pakistan}

\author{Mubasher Jamil}
\email{mjamil@sns.nust.edu.pk}
\affiliation{School of Natural Sciences, National University of Sciences and Technology, H-12, 44000-Islamabad, Pakistan}
\author{Mustapha Azreg-A\"inou}
\email{azreg@baskent.edu.tr}
\affiliation{Ba\c{s}kent University, Engineering Faculty, Ba\u{g}lica Campus, 06790-Ankara, Turkey}

\begin{abstract}
Various studies show that the gravitational collapse of inhomogeneous matter clouds leads to naked singularity formation. We investigate here the spin precession frequency of a test gyroscope attached to a stationary observer in a rotating naked singularity spacetime. In the weak field limit, Lense-Thirring precession for rotating naked singularity and geodetic precession in the asymptotic limit for null naked singularity are found to be equal to that of a Kerr black hole and a Schwarzschild black hole, respectively. In addition, we can distinguish a rotating naked singularity and a Kerr naked singularity for an observer in the equatorial plane using spin precession. To this end, we have found the constraints on the parameters of rotating naked singularity by employing the Monte Carlo Markov Chain simulation and using the observation from five quasi-periodic sources within the relativistic precession model. Our analysis shows that the measurement of spin parameter estimate for GRO J1655-40 is in disagreement with the value found from the continuum-fitting method, while for XTEJ1859+226 and GRS 1915+105, it is inconsistent with spectral analysis results. 
\end{abstract}
\newpage
\maketitle

\section{Introduction}

In modern physics, Einstein's general relativity (GR) theory is a triumph, having withstood numerous tests in both astronomy and astrophysics, including the perihelion precession of Mercury \cite{perihelion}, light deflection by gravity \cite{LightDeflection}, gravitational redshift \cite{pound1960apparent}, Shapiro time delay \cite{shapiro1964fourth}, observations of binary pulsars \cite{pulsars}, detection of gravitational waves \cite{GWs}, and black hole imaging \cite{EHT}. A prediction of general relativity is the existence of black holes, which offer natural testing grounds for gravity in the strong field domain. When the gravitational collapse of a massive star occurs, the formation of a spacetime singularity is inevitable under certain conditions \cite{hawking1967occurrence,hawking1970singularities,hawking2014singularities}. The \textit{Weak Cosmic Censorship Conjecture}, first proposed by Roger Penrose in 1969, states that strong spacetime singularities without horizon are not possible \cite{Roger}. However, this conjecture needs further verification. The weak cosmic censorship conjecture has been tested for several $1+2$, $1+3$, and $1+4$ dimensional black holes \cite{Duztas:2018fbc,Shaymatov:2019del,Shaymatov:2020wtj,Duztas:2020xnl}.  In literature, there are numerous conceivable scenarios in which a visible singularity exists, formed due to the gravitational collapse \cite{10.1093/mnras/stab1186,PhysRevD.102.044037,Mosani_2023}. It can be formed as an equilibrium
end state obtained through the dynamical evolution of a matter cloud starting with regular initial conditions \cite{Joshi_2011}. It can
be demonstrated that the singularity formed during the homogeneous dust collapse is always enclosed within the event horizon. In contrast, any inhomogeneity in the density of the collapsing star along with certain initial conditions leads to the singularity, which can be visible \cite{homo1,homo5,homo6,homo7}. 
There are still unanswered problems regarding the existence of naked singularities in nature and how they differ from black holes in terms of their physical characteristics \cite{joshi_gravitational_2007}.

\par
Both black hole and naked singularity models possess distinct physical properties, leading to unique observational aspects. Researchers have explored these differences extensively, analyzing the observational aspects associated with various naked singularity spacetimes. The properties of an accretion disk around a naked singularity spacetime are discussed in \cite{accretion,Joshi_accretion,Guo_2021_Accretion}. It has been shown that a spacetime without an event horizon and photon sphere can cast a shadow \cite{Dey:2020bgo,Joshi:2020tlq}, and the naked singularity spacetime casts a similar shadow to that of a Schwarzschild black hole in some cases \cite{Shaikh:2018lcc}. In \cite{Joshi:2020tlq}, a static and spherically symmetric null-naked singularity spacetime is described, and a rotating version of null naked singularity (NNS) spacetime has been constructed \cite{patel_light_2022}  using the Newman-Janis algorithm without complexification \cite{azreg2014generating}. By studying the geodesic deviation equation, research indicates that NNS behaves similarly to a Schwarzschild black hole in the asymptotic limit, making the former a mimic of the
latter \cite{Madan2022TidalFE}.  There are also some other naked singularity solutions that have central timelike singularities
, viz Joshi-Malafarina-Narayan (JMN) spacetimes \cite{Joshi_accretion} and Janis-Newman-Winicour (JNW) spacetime \cite{janis_newman}.
\par
The precession of a gyroscope induced by the curvature of spacetime is called \textit{geodetic precession} and can be explained via the \textit{missing inch} model \cite{Overduin:2013tma}. de Sitter proposed in 1916 that the spin axis of an orbiting gyroscope will eventually change direction due to the curvature of local spacetime caused by the gravitational body \cite{deSitter}. Later, Thorne demonstrated how the effect might be understood to originate from two distinct sources: a spin-orbit precession and a space curvature around the central mass \cite{Thorne}. The geodetic precession in the Schwarzschild black hole has been studied in \cite{Hartle:2003yu,SchGeodetic}. A gyroscope
attached to a stationary observer can also precess due to an effect called \textit{Lense-Thirring (LT) precession}, which
causes the dragging of locally inertial frames along the rotating spacetime \cite{Mashhoon:1984fj}. The LT-precession in the strong gravitational field of the Kerr black hole has been discussed in \cite{Chakraborty:BHandNS}. Specifically, it was suggested in \cite{Chakraborty:2016mhx} that one can use a test gyro's spin precession to discern between Kerr naked singularities and black holes. This idea was subsequently expanded upon in modified theories of gravity \cite{Stepanian:2020vwk}. In \cite{Bambhaniya:2021jum} LT precession frequency in slowly rotating naked singularity spacetimes is investigated. Perihelion precession of the particle orbits in naked singularity spacetime has been found and compared with Schwarzschild black hole in \cite{timelike_geodesic}. Schiff suggested in 1959 that test gyroscopes may be employed to measure the wrapping and twisting of spacetime around the Earth \cite{Schiff}. In 2004, NASA launched a satellite, Gravity Probe B, and announced the results in 2011 \cite{Everitt:2011hp}.
\par
Quasi-periodic oscillations (QPOs) are frequently detected within the X-ray power density spectrum of 
X-ray binary systems (XRBs) which comprise a black hole or a neutron star accreting material from a neighboring star. In the vicinity of the compact object, this material becomes heated to such high temperatures that it shines brightly in X-rays. It is observed as relatively narrow peaks \cite{vanderKlis:milliarsec} in the power density spectrum, which offers a method to indirectly chart the flow of accretion, specifically, QPOs. QPOs can be classified into two types: low-frequency QPOs (LFQPOs) with frequency $\lesssim 30$Hz and high-frequency QPOs (HFQPOs) with frequency $\gtrsim 60$Hz \cite{Morgan_1997}, with the former being commonly observed in black hole binaries \cite{2019NewAR..8501524I} and the latter being rare, was first detected by the Rossi X-ray Timing Explorer (RXTE) in black hole systems.  LFQPOs are further classified into types: Type-A, B, and C \cite{LFQPOs}. Three QPOs have been detected simultaneously, with two being HFQPOs for GRO J1655-40 \cite{GRO-HFQPO}. Although the main cause of the production of QPOs is unknown, the frequencies of QPOs can provide crucial information about the mass, spin, and other parameters of the black hole or neutron star. Various models have been proposed to explain the QPOs, viz, relativistic precession model (RPM), resonance models, tidal disruption model, and wrapped disk oscillation model \cite{model1,model2,model3,model4,model5,model6}.
\par
In \cite{Motta:precise-mass}, using the QPOs observation of GRO 1655-40, authors used the RPM to obtain the simultaneous measurements of spin and mass without any prior assumptions, and the mass measurements were found to be consistent with the one obtained from the optical observations \cite{quiescent-meas} along with a well-constrained spin parameter. First proposed in \cite{Stella:Proposed-RPM} to explain QPOs in low mass X-ray binary system (LMXBs) with a neutron star, this model relates the QPOs to the combination of fundamental frequencies (Keplerian (orbital) frequency and epicyclic frequencies) of a test particle which are determined only by the metric. Considering the Kerr black hole, an analytic solution to the RPM system for the case where three QPOs are detected is presented in \cite{Ingram:SolutionsRPM}, thus reducing the computational load. We can also incorporate this model to obtain parameters for twin QPOs. Some other estimates are also reported for spin parameters using the continuum fitting method \cite{Shafee:2005ef} and K$\alpha$ iron line method\cite{IRON-spin} for GRO J1655-40. The use of the continuum-fitting method and the K$\alpha$ iron line together for analysis and measuring the Kerr spin of the same object has been highlighted in \cite{Bambi:combo}. The agreement between the spin measurement of XTEJ1550-564 \cite{Motta:XTE564} and mass measurement of XTE J1859+226 \cite{Motta:XTE226} obtained through RPM to that obtained through the continuum-fitting method  \cite{continum564} and dynamical measurements \cite{Rizo:226}, respectively, strengthens the validity of RPM. The mass and spin of the neutron star have also been obtained via the relativistic precession model \cite{neutron}. Assuming the central compact object to be a superspinning body (or naked singularity) described by the Kerr spacetime with spin parameter $a>M$, it was concluded that within the framework of the epicyclic resonance model, there is a potential way to directly observe and identify the presence of a super-spinning compact object \cite{Kotrlova:kerr-naked}. 
\par
Traditionally, the Kerr metric has been the standard model for describing astrophysical black holes. However, discrepancies in the spin parameter estimates derived from different observational techniques persist. Considering the Kerr spacetime, the measurement of spin parameter reported in \cite{Motta:precise-mass} and \cite{Bambi:testing-rpm} disagrees with those from the continuum fitting method \cite{Shafee:2005ef}. Similarly, it was also argued in \cite{Bambi:testing-rpm} that while GRO J1655-40's data may fit a Kerr black hole, significant deviations continue to exist, as the discrepancy between relativistic precession and continuum fitting persists even without the Kerr assumption. These discrepancies motivate the exploration of alternative rotating spacetimes. Given this context, it would be worth examining whether the RNS model can offer a viable alternative and whether it resolves or reinforces the persistent tension in spin parameter estimates.
\par
With the aforementioned studies and motivation, it will be interesting to explore the behavior of the test gyroscope near rotating naked singularity (RNS) and NNS and determine how it differs from Kerr spacetime. Thus,
one of the main aims is to investigate the geodetic effect in NNS spacetime and general spin precession frequency, including the LT frequency near RNS. We also aim to distinguish between the Kerr naked singularity (described by the Kerr spacetime with spin parameter $a>M$) and RNS. We will explore the behavior of QPOs frequencies near the radius of the innermost stable circular orbit (ISCO) and find the constraints on the parameters of RNS within RPM. In literature, constraints on the parameters are found using different statistical techniques, viz, Monte Carlo Markov Chain (MCMC) \cite{emcee}, $\chi^2$ minimization \cite{Bambi:testing-rpm,jiang_testing_2021,Shaymatov_chi-sqau}, and non-linear square fit \cite{Azreg-Ainou:qpos}. We aim to utilize MCMC for this purpose.
\par
This paper is divided into five sections. In Section II, we will provide a brief introduction to RNS spacetime. In Section III, spin precession frequencies of the test gyroscope are studied in detail. Section IV discusses how we can use the general spin precession frequency to distinguish between a Kerr naked singularity and RNS. Then, we will find the governing equation for ISCO and QPOs frequencies in Section V. Section VI deals with the parameter estimation for RPM using the MCMC. Finally, we will provide a brief conclusion to our results. Throughout the paper, we use the metric signature \((-,+,+,+)\) and units in which $ G = c = 1 $, unless stated otherwise. All Greek indices run from 0 to 3 hereafter, representing $t,r,\theta,\phi$.

\section{Naked Singularity Spacetime}
\label{section-1}
The metric for RNS was derived by the non-complexification procedure modifying the Newman-Janis algorithm~\cite{azreg2014generating} and it is given by~\cite{patel_light_2022}
\begin{align}
\label{2.1}
ds^{2}=-\Big(1-\frac{2f}{\rho^{2}}\Big)dt^{2}-\frac{4af\sin^{2}{\theta}}{\rho^{2}} dtd\phi+\frac{\rho^{2}}{\Delta}dr^{2}+\rho^{2}d\theta^{2}+\frac{\Sigma\sin^{2}{\theta}}{\rho^{2}} d\phi^{2},
\end{align}
where
$$\rho^{2}=r^{2}+a^{2}\cos^{2}{\theta},\ \ 
\Delta=\frac{r^{4}}{(r+M)^{2}}+a^{2},\ \  
f=\frac{r^{2}M^{2}+2Mr^{3}}{2(M+r)^{2}},\ \ 
\Sigma=(r^{2}+a^{2})^{2}-a^{2}\Delta\sin^{2}{\theta}.$$
\par
Here, $a$ is the spin parameter of RNS. The line element (\ref{2.1}) represents a solution to the Einstein field equations with matter sources and satisfies the energy conditions as described in Eqs. (27)- (29) of~\cite{patel_light_2022}
The event horizon does not exist in this spacetime. To demonstrate this non-existence statement, we can check for the nature of roots of the equation $g^{rr}=0$, given by
\begin{align}
r^4+a^2(r+M)^2=0,
\end{align}
which admits no real solution since its left-hand side is manifestly positive.
The absence of real roots suggests the absence of an event horizon, exhibiting geometrical characteristics contrary to those of a black hole. Similarly, we can also check for the equation of ergoregion by taking $g_{tt}=0$ and deduce that ergoregion is also absent for RNS. Substituting $a=0$ in (\ref{2.1}) reduces it to a static and spherically symmetric NNS spacetime given by
\begin{equation}
\label{2.5}
ds^2 = -\left(1 + \frac{M}{r}\right)^{-2} dt^2 + \left(1 + \frac{M}{r}\right)^{2} dr^2 + r^2 \left( d\theta^2 + \sin^2\theta d\phi^2 \right),
\end{equation}
where $M$ denotes the ADM mass. The event horizon equation $g^{rr}=0$ for the above spacetime (\ref{2.5}), implies that the event horizon does not exist as well, however, a curvature singularity occurs at $r=0$ in the equatorial plane (taking $\theta=\pi/2$) which can be checked using the curvature invariants for RNS and NNS, as demonstrated in \cite{patel_light_2022} and \cite{Joshi:2020tlq} respectively. 
\par
Causality violations and closed timelike curves (CTCs) in RNS are possible in a region where $g_{\phi\phi}=\frac{\Sigma\sin^{2}{\theta}}{\rho^{2}}<0$ \cite{azreg2014generating}. Since $\frac{\sin^{2}{\theta}}{\rho^{2}}>0$, $g_{\phi\phi}$ and $\Sigma$ have the same sign. The latter is brought to the form
\begin{equation}
\Sigma =r^4+\frac{a^2 r^2 \left(2 M^2+4 M r+r^2\right)}{(M+r)^2}+\Big[a^2+\frac{r^4}{(M+r)^2}\Big] a^2 \cos^2\theta\,,
\end{equation}
which is manifestly positive for all $r\geq 0$ and $0\leq\theta\leq\pi$. Thus, no causality violations occur in the physical region $r\geq 0$ and arbitrary $\theta$ (outside the singularity). 
\section{General Spin Precession}
\label{section-2}
\begin{figure}
\centering
\begin{subfigure}{0.5\textwidth}
	\centering
	\includegraphics[width=\linewidth]{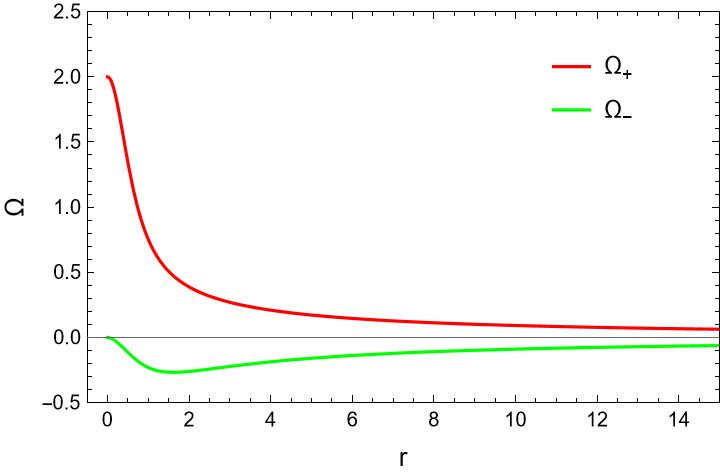}
	\caption{$a=0.5$}
\end{subfigure}%
\begin{subfigure}{0.5\textwidth}
	\centering
	\includegraphics[width=\linewidth]{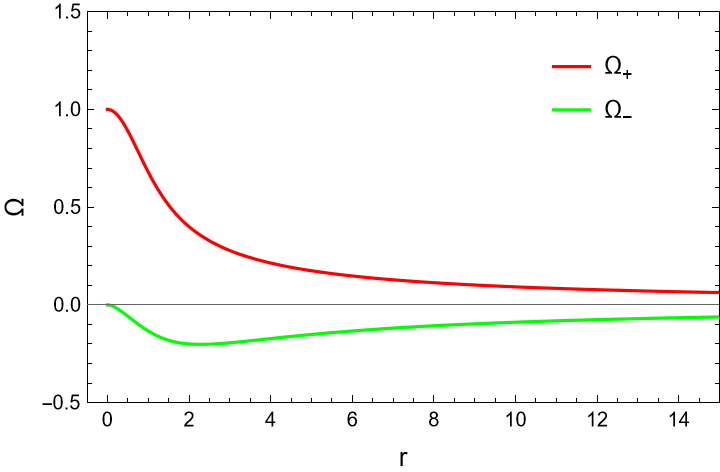}
	\caption{$a=1.0$}
\end{subfigure}
\caption{Roots of angular velocities $\Omega_{+}|_{\theta=\frac{\pi}{2}}$ and $\Omega_{-}|_{\theta=\frac{\pi}{2}}$ by taking $M=1$. It can be seen that both curves do not intersect each other at any point. }
\label{fig:angular}
\end{figure}

In this section, we address the spin precession frequency of the test gyroscope attached to a stationary observer in RNS spacetime. Let us consider a test gyroscope in a stationary spacetime attached to an observer at rest. The observer moves along an integral curve $\gamma(\tau)$ of the timelike Killing vector $K=\partial_{t}+\Omega\partial_{\phi}$ (also known as the Killing trajectory) of spacetime to stay at rest in the stationary spacetime \cite{Nstraumann}. So, its velocity vector is $\underline{u}=(-<K,K>)^{-\frac{1}{2}}K$. Following the steps in \cite{AzregAnou2019GyroscopePF,Nstraumann}, the one form of the general spin precession frequency of a test stationary gyroscope becomes
\begin{align*}
\tilde{\Omega}_{p}=\frac{1}{2K^2}\star(\tilde{K} {\wedge} d \tilde{K}),
\end{align*}
where $\star$ is the Hodge star operator, $\wedge$ is the wedge product, and $\tilde{K}$ is the corresponding co-vector of $K$ given by $\tilde{K}=g_{t\mu}dx^{\mu}+\Omega g_{\phi\beta}dx^{\beta}$. More details on the derivation of this equation have been given in Appendix of Ref.~\cite{AzregAnou2019GyroscopePF} and in~\cite{Straumann}. 
As shown in Appendix A, we bring this equation of the general spin precession of the test gyroscope attached to a timelike stationary observer in a stationary axi-symmetric spacetime to the form~\cite{Chakraborty:2016mhx,AzregAnou2019GyroscopePF,Chakraborty_2017,Wu:2023wld}
\begin{align}
\label{den}
\Vec{\Omega}_\text{p} &=\dfrac{W\hat{r}+Z\hat{\theta}}{2\sqrt{-g}(g_{tt}+2\Omega g_{t\phi}+\Omega^{2} g_{\phi\phi})},
\end{align}
where
\begin{eqnarray*}
W&=&-\sqrt{g_{rr}} \left[\left(g_{tt}g_{t\phi,\theta} - g_{t\phi} g_{tt,\theta}\right) + \Omega \left(g_{tt}g_{\phi\phi,\theta} - g_{\phi\phi} g_{tt,\theta}\right) + \Omega^{2} \left(g_{t\phi} g_{\phi\phi,\theta} - g_{\phi\phi} g_{t\phi,\theta}\right)\right],\nonumber\\
Z&=&\sqrt{g_{\theta\theta}} \left[\left(g_{tt}g_{t\phi,r} - g_{t\phi} g_{tt,r}\right) + \Omega \left(g_{tt}g_{\phi\phi,r} - g_{\phi\phi} g_{tt,r}\right) + \Omega^{2} \left(g_{t\phi}g_{\phi\phi,r} - g_{\phi\phi} g_{t\phi,r}\right)\right].
\end{eqnarray*}
Here, $\hat{r}$ and $\hat{\theta}$ represent the unit vectors in the radial and polar directions, respectively, and $g$ denotes the determinant of the metric. For timelike observers $u^{\mu}=u^{t}(1, 0, 0,\Omega)$, the above expression (\ref{den}) holds true for a limited range of $\Omega$~\cite{rizwan_distinguishing_2018}, which will be given shortly later [see Eq.~\eqref{const}]. The general spin precession frequency for the metric (\ref{2.1}) becomes
\begin{footnotesize}
\begin{align}
	\label{3.1}
	\Vec{\Omega}_\text{p} =& \dfrac{X\cos\theta\sqrt{\Delta} \hat{r} + Y \sin{\theta} \hat{\theta}}{\rho^3 \bigl[2 a M r^2 \sin^2\theta  (M+2 r) \Omega +\left( a^2 \cos^2\theta  (M+r)^2+ r^4 \right)  - \Omega^2\sin^2\theta \left\{(a^2+r^2)^2 (M+r)^2 - a^2 \sin^2\theta  (a^2 (M+r)^2+r^4)\right\}\bigr]},
\end{align}
\end{footnotesize}
where
{\small
\begin{align*}
	X=&a M r^2 (M+2 r)-\frac{1}{8} \Omega  \biggl[a^4 \cos 4 \theta  (M+r)^2+3 a^4 (M+r)^2+8 a^2 r^2 \bigl(2 M^2+4 M r+r^2\bigr)\nonumber\\
	&+4 a^2 \cos 2 \theta  \left\{a^2 (M+r)^2+2 r^4\right\}+8 r^4 (M+r)^2\biggr]+ \Omega ^2 a^3 M r^2 \sin ^4\theta  (M+2 r),\\
	Y=&-\frac{a M r}{(M+r)} \biggl[a^2 \cos ^2\theta  \bigl(M^2+3 M r+r^2\bigr)-r^4\biggr]+ \frac{r \Omega}{2 (M+r)}\biggl[a^2 \cos ^2\theta  \biggl\{a^2 \bigl(4 M^3+12 M^2 r+10 M r^2+3 r^3\bigr)\nonumber\\
	&+4 r^2 (M+r)^3\biggr\}+r^2 \biggl\{-a^4 \sin ^2\theta  \cos 2 \theta  (2 M+r)-a^4 (2 M+r)-4 a^2 M r^2+2 r^5\biggr\}\biggr]-\frac{a M r \sin ^2\theta\Omega^2}{2 (M+r)} \nonumber\\
	&\times \biggl[a^4 \bigl(M^2+3 M r+r^2\bigr)+a^2 \cos 2 \theta  \left\{a^2 \bigl(M^2+3 M r+r^2\bigr)-r^4\right\}-3 a^2 r^4-2 r^4 \bigl(M^2+3 M r+3 r^2\bigr) \biggr].
\end{align*}}
As the equation for ergoregion has no real roots for RNS, the expression (\ref{3.1}) is valid everywhere outside the singularity. In the above expression, the allowed values of \( \Omega \) at any fixed \( (r, \theta) \) for timelike observers are constrained by
\begin{align}\label{const}
\Omega_{-}(r,\theta)<\Omega(r,\theta)<\Omega_{+}(r,\theta),
\end{align}
with
\begin{eqnarray}
\label{3.3}
\Omega_{\pm}&=& \frac{2af\sin{\theta}\pm\sqrt{4a^{2}f^{2}\sin^{2}{\theta}+\Sigma(\rho^{2}-2f)}}{\Sigma\sin{\theta}},\nonumber\\&=&\frac{a M r^2 (M +2 r) \sqrt{1-u^2}\pm (r+M) (r^2+a^2 u^2) \sqrt{r^4+a^2 (M+r)^2}}{\sqrt{1-u^2} [(M+r)^2 (r^4
	+a^4u^2)+a^2 r^2 \{2 M^2+4 M r+r^2 (1+u^2)\}]},
\end{eqnarray}
being the roots of the denominator in~(\ref{den}) and $u=\cos\theta$. These expressions indicate that $\Omega_{+}$ and $\Omega_{-}$ become undetermined near the singularity $(r = 0, \theta= \frac{\pi}{2})$. However, we can show that $\lim_{(r\to 0^+,\,u\to 0)} \Omega_{+}=1/a$ and that $\lim_{(r\to 0^+,\,u\to 0)} \Omega_{-}$, which remains finite, does not exist (details are given in the Appendix B). The limit $\lim_{(r\to 0^+,\,u\to 0)} \Omega_{-}$ depends on the shape of the path $r=g(u)$, with $g(0)=0$, through the singularity, that the timelike observer borrows and its value is always finite regardless of the shape of $r=g(u)$. This simply means that the lower limit of the domain of convergence of the spin precession~(\ref{den}), as the observer approaches the singularity, depends on the path she follows. If the motion of the timelike observer is confined in the $u=0$ ($\theta=\pi/2$) plane, $\Omega_{\pm}$ reduces to
\begin{align}
\Omega_{\pm}|_{\theta=\frac{\pi}{2}}=  \frac{a M (M+2 r) \pm (M+r) \sqrt{ \left(a^2 (M+r)^2+r^4\right)}}{a^2  \left(2 M^2+4 M r+r^2\right)+r^2 (M+r)^2},
\end{align}
where the r.h.s has existing limits as $r\to 0^+$, yielding $0<\Omega(r,\pi/2)<1/a$~(\ref{const}) in the vicinity of the singularity if $a>0$ or $1/a<\Omega(r,\pi/2)<0$ if $a<0$. From now on, we restrict ourselves to the case $a>0$. The behavior of $\Omega_{\pm}|_{\theta=\frac{\pi}{2}}$ at $a=0.5$ and $a=1.0$ is depicted respectively in panels (a) and (b) of Fig.\ref{fig:angular}. We see that the domain of convergence of the spin precession~(\ref{den}) enlarges as one approaches the singularity. Roots of angular velocity do not intersect each other, unlike the Kerr naked singularity, where the curves meet each other near the singularity, as shown in \cite{Chakraborty:2016mhx}.
\par
\begin{figure}
\centering
\begin{subfigure}{0.33\textwidth}
	\centering
	\includegraphics[width=\linewidth]{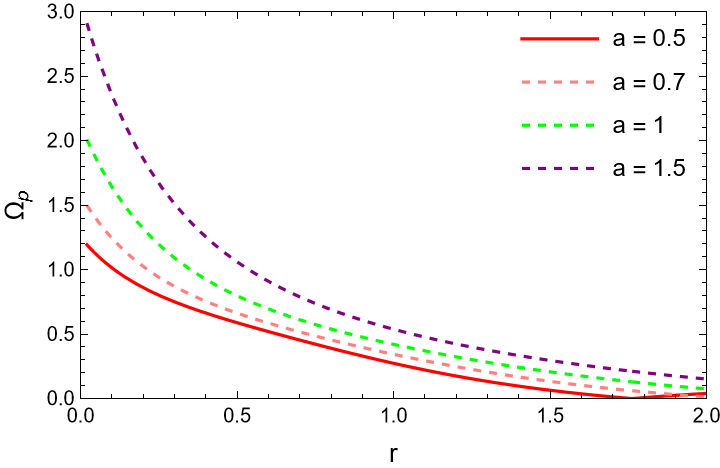}
	\caption{$\theta=\frac{\pi}{2},~k=0.2$}
\end{subfigure}%
\begin{subfigure}{0.33\textwidth}
	\centering
	\includegraphics[width=\linewidth]{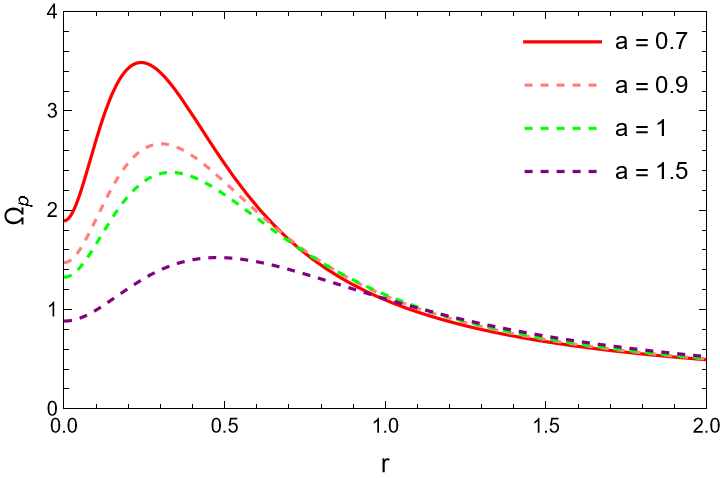}
	\caption{$\theta=\frac{\pi}{4},~k=0.2$}
\end{subfigure}%
\begin{subfigure}{0.33\linewidth}
	\centering
	\includegraphics[width=\linewidth]{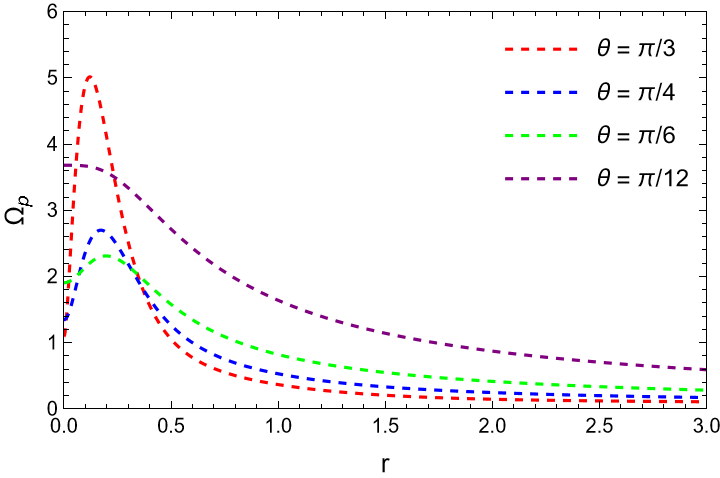}
	\caption{$a=0.5,~k=0.3$}
\end{subfigure}
\begin{subfigure}{0.33\linewidth}
	\centering
	\includegraphics[width=\linewidth]{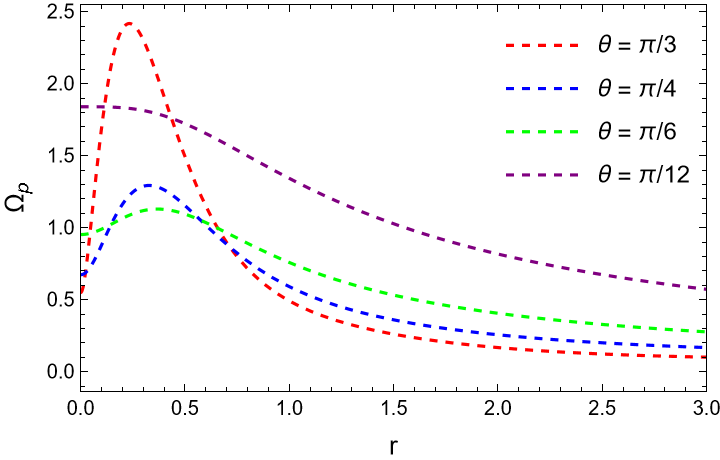}
	\caption{$a=1,~ k=0.3$}
\end{subfigure}%
\begin{subfigure}{0.33\linewidth}
	\centering
	\includegraphics[width=\linewidth]{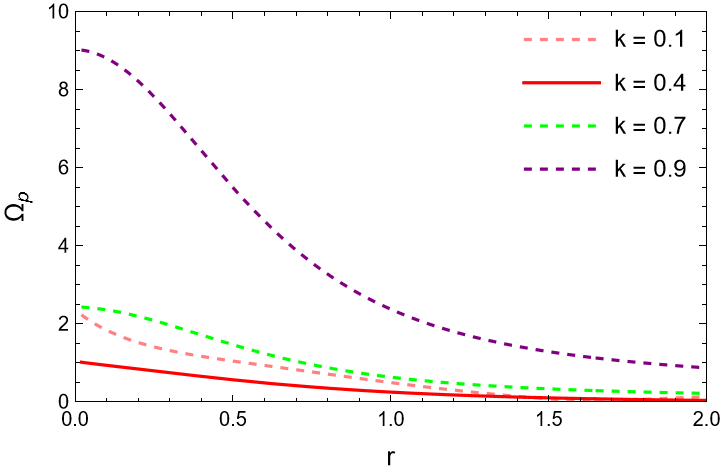}
	\caption{$a=0.5,~\theta=\frac{\pi}{2}$}
\end{subfigure}%
\begin{subfigure}{0.33\linewidth}
	\centering
	\includegraphics[width=\linewidth]{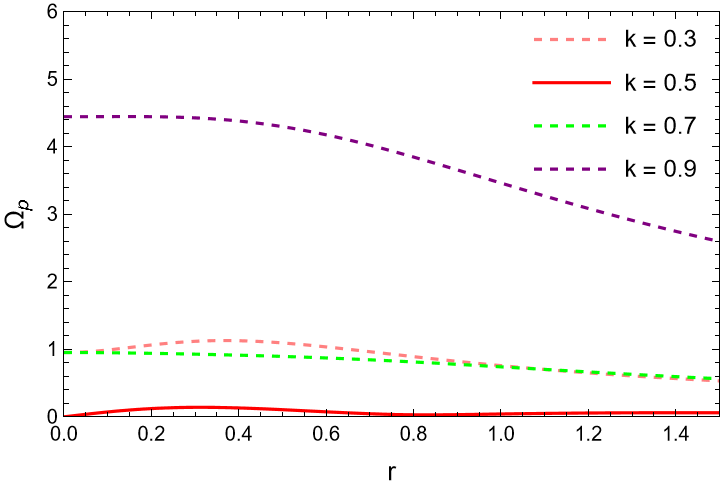}
	\caption{$a=1, ~\theta=\frac{\pi}{6}$}
\end{subfigure}
\caption{General spin precession frequency for a RNS for $M=1$ and different values of the spin parameter \( a \), dimensionless parameter $k$ and the angle \( \theta \). It is observed that precession frequency remains finite for all observers except at the singularity. For an observer situated at \( \theta = \pi/2 \), the frequency remains finite throughout, but diverges at the singularity itself. The event horizon and ergoregion are absent everywhere, distinguishing this scenario from the typical Kerr black hole and Kerr naked singularity case.}
\label{Fig.1}
\end{figure}
\begin{figure}
\centering
\begin{subfigure}{0.33\linewidth}
	\centering
	\includegraphics[width=\linewidth]{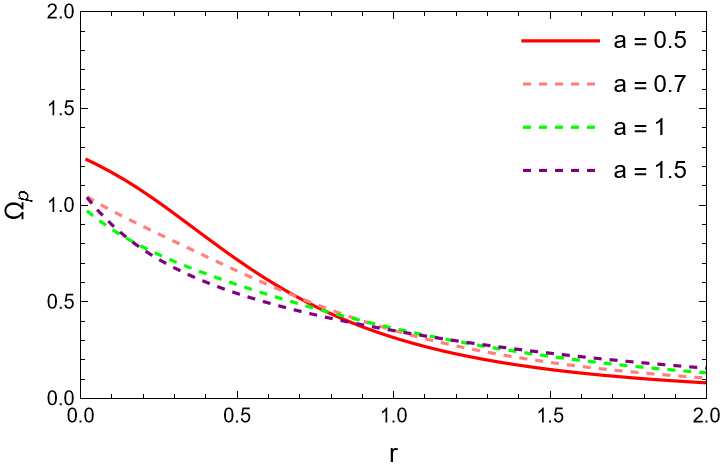}
	\caption{$\theta=\frac{\pi}{2}$}
\end{subfigure}%
\begin{subfigure}{0.33\linewidth}
	\centering
	\includegraphics[width=\linewidth]{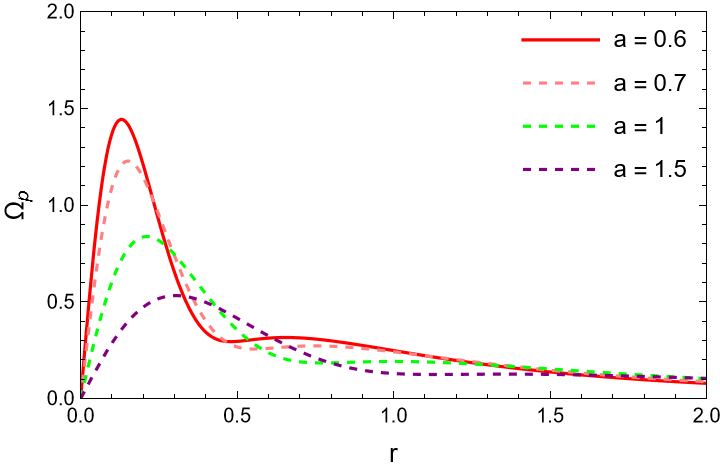}
	\caption{$\theta=\frac{\pi}{3}$}
\end{subfigure}%
\begin{subfigure}{0.33\linewidth}
	\centering
	\includegraphics[width=\linewidth]{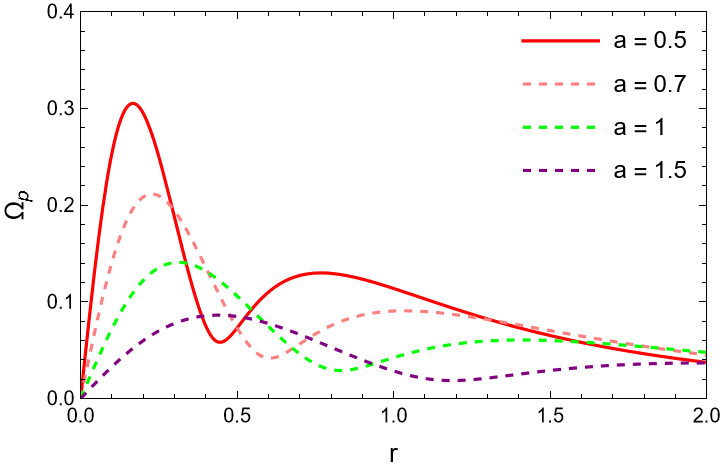}
	\caption{$\theta=\frac{\pi}{6}$}
\end{subfigure}
\caption{Magnitude of the general spin precession plotted for a ZAMO by taking $(M=1)$. It can be seen that the precession frequency remains finite everywhere for all observers except at the singularity.}
\label{Fig.2}
\end{figure}

We can introduce the dimensionless parameter $k$ to write angular velocity $\Omega$ in terms of $\Omega_{\pm}$ so that it covers the entire range of $\Omega$ from $\Omega_{+}$ to $\Omega_{-}$
\begin{align*}
\Omega=k\Omega_{+}+(1-k)\Omega_{-},
\end{align*}
where $0<k<1$. Hence, we obtain
\begin{align}
\Omega=\frac{2af\sin{\theta}+(2k-1)\sqrt{4a^{2}f^{2}\sin^{2}{\theta}+\Sigma(\rho^{2}-2f)}}{\Sigma\sin{\theta}}.
\end{align}
Now we can simplify the denominator of (\ref{3.1}) to obtain a more decent expression for the magnitude of the general spin precession frequency in terms of $k$ as
\begin{align}
\Omega_\text{p}= \frac{\left(a^2+r^2\right)^2 (M+r)^2-a^2 \sin ^2\theta \left(a^2 (M+r)^2+r^4\right)}{4 k (1-k) (M+r)^2 \left(a^2 (M+r)^2+r^4\right) \rho^{7}} \sqrt{X^2|\Delta| \cos^{2}\theta + Y^2 \sin^{2}{\theta}} .
\end{align}
The denominator of the above equation vanishes at the singularity. For the zero angular momentum observer (ZAMO), we take $k=0.5$ so the angular velocity becomes

\begin{align}
\Omega|_{k=0.5}=\frac{-g_{t\phi}}{{g_{\phi\phi}}}=\frac{2af}{\Sigma}.
\end{align}
The above expression is the angular velocity of freely falling particles (initially at rest at spatial infinity) as they fall into the central object’s influence, illustrating the frame-dragging effect \cite{azreg2014generating}. The gyroscopes attached to ZAMO are non-rotating with respect to the local geometry, and stationary observers regard both $+\phi$ and $-\phi$ as the same \cite{rizwan_distinguishing_2018}. Thus, it is interesting to study how the precession of a gyroscope attached to a ZAMO behaves. We can simplify the expressions for $X$ and $Y$ for ZAMO as
\begin{align}
X|_{k=0.5}&=\frac{1}{C^2} \biggl[a^3 M r^2 \sin ^2\theta (M+r)^2 (M+2 r) \left(a^2 (M+r)^2+r^4\right) \left(a^2 \cos (2 \theta )+a^2+2 r^2\right)^2\biggr],
\end{align}
\begin{align}
Y|_{k=0.5} &= \frac{-1}{2 C^2} \biggl[a M r (M+r)\left(a^2 (M+r)^2 + r^4\right) \left(a^2 \cos (2 \theta) + a^2 + 2 r^2\right)^2 \biggl\{ a^4 \left(M^2 + 3 M r + r^2\right) \nonumber \\
&\quad + a^2 \cos (2 \theta) \bigl\{ a^2 \left(M^2 + 3 M r + r^2\right) - r^4 \bigr\} - 3 a^2 r^4 - 2 r^4 \left(M^2 + 3 M r + 3 r^2\right) \biggr\} \biggr],
\end{align}

where
\begin{align*}
C= a^4 (M+r)^2+a^2 \cos (2 \theta ) \left(a^2 (M+r)^2+r^4\right)+a^2 r^2 (2 M+r) (2 M+3 r)+2 r^4 (M+r)^2.
\end{align*}
In Fig.~\ref{Fig.1}, plots of the general spin precession frequency (\ref{3.1}) are given for different cases. General spin precession remains finite for all observers. However, it becomes undetermined for an observer in the equatorial plane at $r=0$. This behavior makes RNS different from the Kerr black hole and Kerr naked singularity, which will be discussed in the next section. For the Kerr black hole, the spin precession diverges at the event horizon for all observers \cite{Chakraborty:2016mhx}. In Fig.~\ref{Fig.2}, we have plotted the general spin precession for ZAMO. Similar to all other cases of $k$, for $k = 0.5$, $\Omega_{p}$ remains finite. For the ZAMO observer in the equatorial plane, it diverges at $r=0$.

\subsection{Lense-Thirring precession frequency}
\begin{figure}
\centering
\begin{subfigure}{0.45\linewidth}
	\centering
	\includegraphics[width=\linewidth]{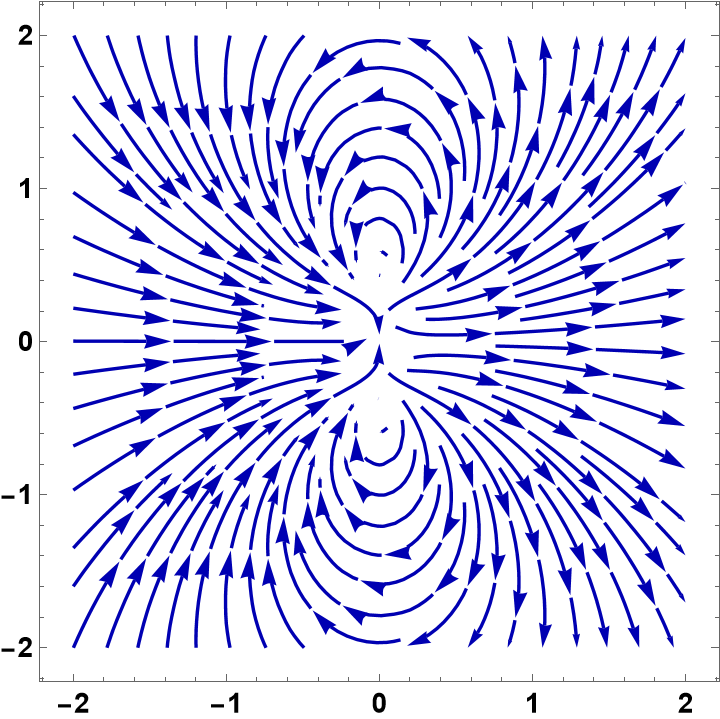}
	\caption{$M=1,\,a=0.5$}
\end{subfigure}%
\begin{subfigure}{0.45\linewidth}
	\centering
	\includegraphics[width=\linewidth]{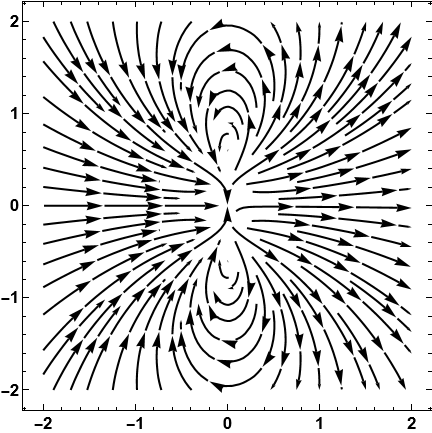}
	\caption{$M=2.5,\,a=0.5$}
\end{subfigure}
\caption{LT precession frequency vector fields for RNS, plotted in Cartesian coordinates corresponding to $(r,\theta)$. The field lines show that for RNS, the LT precession is finite up to the region around the origin $(0,0)$, which corresponds to the location of the singularity. The vectors form circular or elliptical loops, demonstrating the nature of frame dragging caused by the RNS.}
\label{Fig.3}
\end{figure}

\begin{figure}
\centering
\begin{subfigure}{0.33\linewidth}
	\centering
	\includegraphics[width=\linewidth]{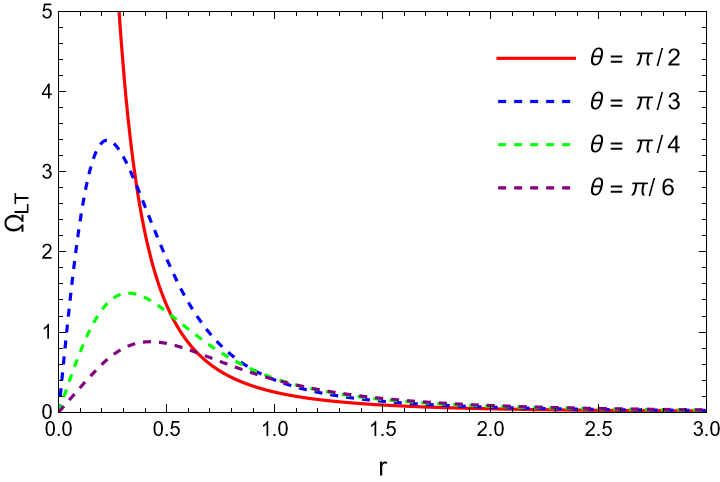}
	\caption{$a=0.5$}
\end{subfigure}%
\begin{subfigure}{0.33\linewidth}
	\centering
	\includegraphics[width=\linewidth]{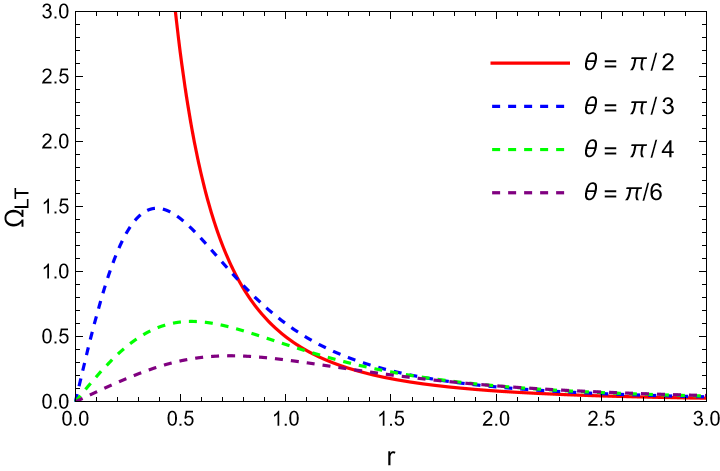}
	\caption{$a=1.0$}
\end{subfigure}%
\begin{subfigure}{0.33\linewidth}
	\centering
	\includegraphics[width=\linewidth]{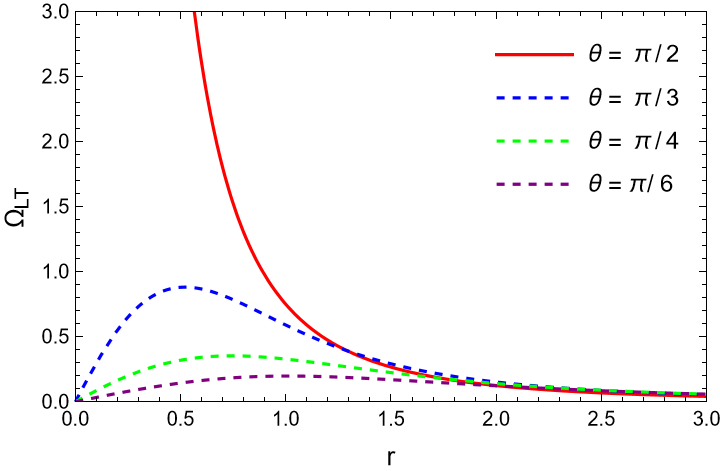}
	\caption{$a=1.5$}
\end{subfigure}%
\caption{Magnitude of LT precession frequency at $M=1$. It remains finite for all observers but diverges for observers with $\theta=\frac{\pi}{2}$ as it approaches the $r \rightarrow 0$. The magnitude of LT precession frequency attains a certain peak for all observers except the one in the equatorial plane. This peak decreases with increasing $a$ and increases with increasing $\theta$.}
\label{Fig.5}
\end{figure}

In Eq.~(\ref{3.1}), both spacetime rotation (LT precession) and curvature (geodetic precession) have an impact on the precession. When angular velocity vanishes, i.e., $\Omega=0$, (\ref{3.1}) reduces to the LT precession frequency \cite{Chakraborty:2016mhx}, meaning that the gyroscope is attached to a static observer. Thus we obtain
\begin{align}
\label{3.11}
\Vec{\Omega}_\text{LT}=\dfrac{(F\sqrt{\Delta}\cos{\theta})\hat{r}+(H\sin{\theta})\hat{\theta}}{\rho ^{3} \bigl[a^2 \cos ^2\theta (M+r)^2+r^4\bigr]},
\end{align}
where
\begin{align*}
F=a M r^2 (M+2 r),\ \ 
H=-\frac{a M r \bigl[a^2 \cos ^2\theta \left(M^2+3 M r+r^2\right)-r^4\bigr]}{M+r}.
\end{align*}
(\ref{3.11}) is the LT precession frequency vector of a test gyro solely due to the rotation of RNS (frame-dragging effect), and it vanishes if $a=0$, as $a$ is a common factor in $F$ and $H$. This LT precession frequency vector vanishes identically at $r=0$ for any $\theta\neq\pi/2$. Vector field plots for LT precession frequency are plotted in the Cartesian plane in Fig.~\ref {Fig.3} corresponding to $(r,\theta)$ for different values of $a$ and $M$. It can be seen that it remains regular
in the whole region except at the singularity $(r=0$ and $\theta=\frac{\pi}{2})$. The magnitude of $\Vec{\Omega}_\text{LT}$ is given by
\begin{align}
\label{3.12}
\Omega_\text{LT}=\dfrac{\sqrt{F^2\Delta\cos^{2}{\theta}+H^2\sin^{2}{\theta}}}{\rho ^{3} \bigl[a^2 \cos ^2\theta (M+r)^2+r^4\bigr]},
\end{align}
which vanishes identically at $r=0$ for any $\theta\neq\pi/2$, while the corresponding expression for Kerr naked singularity behaves as $M\sin\theta /(a^2\cos^3\theta)$ at $r=0$.
On the plane $\theta = \frac{\pi}{2}$, (\ref{3.12}) reduces to 
\begin{align}
\Omega_\text{LT}|_{\theta \rightarrow \frac{\pi}{2}}=\frac{a M}{r^{2}(M+r)}.
\end{align}
Thus, when approaching the singularity from within the $(\theta =\pi/2)$-plane, $\Omega_\text{LT}$ diverges. This shows that $\lim_{(r\to 0,\,\theta\to\pi/2)}$ depends on the path the gyroscope follows to reach $r=0$. The magnitude of the LT precession is plotted in Fig.~\ref{Fig.5}. It can be seen that the LT precession remains finite everywhere except for the observer in the equatorial plane, where it blows up as $r\rightarrow0$. Further, $\Omega_{\text{LT}}$ decreases as $r$ increases and has a peak. The peak increases with increasing angle while the spin parameter has the opposite influence, similar to the Kerr case \cite{Chakraborty:BHandNS}. 
\par
The LT precession vector in a rotating spacetime, without any weak gravity approximation, is given by (\ref{3.11}). In the weak-field limit $(r \gg M)$, (\ref{3.11}) reduces to
\begin{equation}
\Vec{\Omega}_{\text{LT}}(\text{weak}) = \frac{J}{r^3} \left[2 \cos\theta \hat{r} + \sin\theta \hat{\theta}\right],
\end{equation}
where $\theta$ is the colatitude and $J=aM$. This equation bears resemblance with equation $(44)$ of \cite{Chakraborty_2014}, which is the LT precession frequency of Kerr spacetime in the weak field limit $(r\gg M)$. This shows that in weak field limits, RNS and Kerr spacetimes cannot be distinguished. 
\par
Furthermore, in the same limit, the metric (\ref{2.1}) also reduces to the weak-field approximation of the Kerr metric (see Appendix C for details), confirming that both the spacetime and the physical effects, such as Lense-Thirring precession, behave identically in the weak-field regime.

\subsection{Geodetic precession} 
The LT precession vanishes for the line element (\ref{2.5}). This allows us to isolate the effects of spacetime curvature on the precession of the gyroscope in the absence of the frame-dragging effect. The spin vector of a test body with spin (gyroscope) can be used to probe the geometry of spacetime. Suppose that the spin of the test body is described by the 4-vector $\boldsymbol{S}(\tau)$ along the geodesic. In the rest frame, this vector has no timelike component, so that we can write the orthogonality condition
\begin{align}
\label{3.15}
S \cdot u = g_{\mu\beta}S^{\mu}u^{\beta} = 0.
\end{align} 
As the gyroscope moves along a timelike geodesic curve, components of its 4-velocity $u^{\mu}$ satisfy the geodesic equation, and to ensure the conservation of inner products at all points, components of $\boldsymbol{S}(\tau)$ must satisfy \cite{zhu2024observational}
\begin{align}
\label{3.16}
\frac{dS^\mu}{d\tau} + \Gamma^{\mu}_{\beta\gamma}S^\beta u^\gamma = 0,
\end{align}
which effectively says that $S^\mu$ is parallel transported along $u^\mu$, where $\tau$ is the proper time along the worldline, $\Gamma^{\mu}_{\beta\gamma}$ are the Christoffel symbols. Without loss of generality, we consider the particle in the equatorial plane $(\theta=\pi/2)$. For a test body in a circular equatorial motion, we have $u^r = u^{\theta} = 0$, so the components of 4-velocity are given by $u^{\mu} = u^t(1, 0, 0, \Omega)$, where $u^t = \sqrt{\frac{(r+M)^3}{r^3}}$ and $\Omega^2 = \frac{M}{(r+M)^3}$ are both constants. Now, the orthogonality condition  leads to
\begin{align}
S^t = (r+M)^2 \Omega S^{\phi} \label{3.17}.
\end{align} 
Putting $\mu= t,r,\theta,\phi$ in (\ref{3.16}) and using non-zero Christoffel symbols, the system of equations becomes
\begin{align*}
\frac{dS^t}{d\tau} +\frac{M}{r(r+M)} S^r u^t = 0, \quad \frac{dS^r}{d\tau} - \frac{r^4 \Omega u^t}{(M+r)^3} S^{\phi} = 0, \quad \frac{dS^{\theta}}{d\tau} = 0, \quad \frac{dS^{\phi}}{d\tau} + \frac{\Omega}{r} S^r u^t = 0.
\end{align*}
It is simple to demonstrate that the first and fourth equations are identical by using (\ref{3.17}) to eliminate $S^t$ from the first equation of the system. It is more suitable to convert the $\tau$ derivatives to $ t$ derivatives using $u^t = dt/d\tau$, and the system of equation reduces to
\begin{align}
\frac{dS^r}{dt} - \frac{r^4 \Omega }{(M+r)^3} S^{\phi} =0, \quad
\frac{dS^{\theta}}{dt} = 0,\quad
\frac{dS^{\phi}}{dt} + \frac{\Omega}{r} S^r = 0 .
\end{align}
Combining the first and last equations yields
\begin{align*}
S^{\phi} (t)=A \cos{(\Omega^{'} t)}+ B \sin{(\Omega^{'} t)},
\end{align*}
where $A$ and $B$ are constants to be determined and $\Omega^{'}=\Omega\sqrt{\frac{r^3}{(M+r)^3}}$. If the initial spin direction is assumed to be radial, then $S^{\theta} (0)=S^{\phi}(0)=0$, which implies $A=0$ and the corresponding solution can be easily obtained after determining $B$. Hence, we obtain the components of the 4-spin as
\begin{align}
S^\mu = S^r (0) \left(-\frac{(M+r)^2 \Omega^{2}}{\Omega^{'}r} \sin{(\Omega^{'} t)},\cos{(\Omega^{'} t)}, 0, -\frac{\Omega }{\Omega^{'} r} \sin{(\Omega^{'} t)} \right).
\end{align}
This solution demonstrates how the spatial components of the spin vector rotate with respect to the radial direction. The radial direction itself rotates in the $+\phi$ at a coordinate angular speed $\Omega$. At $t=0$,
\begin{align*}
S^\mu|_{t=0} = (0,S^r(0),0,0).
\end{align*}
As one revolution is completed in $T=\frac{2\pi}{\Omega}$, the spatial spin vector is rotated by an angle $2\pi+ \Delta\phi_{\text{geodetic}}$ where $\Delta\phi_{\text{geodetic}}=2\pi\Big(1-\frac{\Omega^{'}}{\Omega}\Big)$ which leads to
\begin{align}
\label{3.23}
\Delta\phi_{\text{geodetic}} &= 2\pi\left(1-\sqrt{\frac{r^3}{(M+r)^3}}\right),
\end{align}
which is manifestly different from the Schwarzschild counterpart $\Delta\phi_{\text{geodetic}}^{(\text{Sch})} = 2\pi\left(1-\sqrt{1-\frac{3M}{r}}\right)$. However, both expressions reduce to
\begin{align}
\label{3.24}
\Delta\phi_{\text{geodetic}, \odot}^{(\text{NNS})} &= \frac{3\pi G M}{rc^2},
\end{align}
in the limit of small values of $M/r$ typical in the solar system scale.
Thus, we conclude that in the solar system, geodetic precession calculated from the static spherically symmetric black hole and naked singularity matches the prediction of GR~\cite{Hartle:2003yu}. 

To the second order in $M^2/r^2$, the two expressions $\Delta\phi_{\text{geodetic}}$ and $\Delta\phi_{\text{geodetic}}^{(\text{Sch})}$ are quite different, yielding $-15\pi M^2/(4r^2)$ and $9\pi M^2/(4r^2)$, respectively. Numerically, this second-order correction is of the order of $10^{-6}$ mas per year, which is much smaller than the 1 $\sigma$ error provided by the Gravity Probe B (GPB) \cite{Everitt:2011hp}. When taking into account the GPB at an altitude of 642 km with an orbital period of 97.65 min, the geodetic drift rate is \cite{Everitt:2011hp}
\begin{align}
\Delta\phi_{\text{geodetic}}^{(\text{GPB})}  = -6601.8\pm 18.3 \text{ mas per year}.
\end{align}
With nowadays experimental tools, it is not possible to distinguish the static spherically symmetric black hole and a naked singularity.

\section{Distinguishing a Kerr Naked Singularity from a Rotating Naked Singularity }
\label{section-3}
The Kerr metric in Boyer-Lindquist coordinates $(t, r, \theta, \phi)$ is given by:
\begin{equation}
\begin{aligned}
	\label{4.1}
	ds^2 = & -\left(1 - \frac{2Mr}{(r^2 + a^2 \cos^2\theta)}\right) dt^2 - \frac{4Mar\sin^2\theta}{(r^2 + a^2 \cos^2\theta)} dt \, d\phi + \frac{(r^2 + a^2 \cos^2\theta)}{(r^2 - 2Mr + a^2)} dr^2 + (r^2 + a^2 \cos^2\theta) \, d\theta^2 \\
	& + \left(r^2 + a^2 + \frac{2Ma^2r\sin^2\theta}{(r^2 + a^2 \cos^2\theta)}\right) \sin^2\theta \, d\phi^2,
\end{aligned}
\end{equation}
Here, $M$ is the mass of the black hole and $a$ is its spin parameter. If the spin parameter \(a\) (the angular momentum per unit mass) satisfies the condition \(M \geq a\), the Kerr solution represents a black hole, and the Kerr singularity is contained within the event horizon. However, if \(M < a\), the event horizon disappears, leading to a naked singularity. Spin precession frequencies for spacetime described by the Kerr metric have been studied extensively in \cite{Chakraborty:2016mhx,Chakraborty:BHandNS,Chakraborty_2014}. It was also demonstrated in \cite{Chakraborty:BHandNS} that in the case of a Kerr naked singularity, the ergoregion exists for the ranges $\frac{\pi}{3} \leq \theta_e \leq \frac{\pi}{2}$ and $-\frac{\pi}{3} \geq \theta_e \geq -\frac{\pi}{2}$ for $a/M=2$. The general spin precession frequency of a Kerr naked singularity blows up for $\theta=\frac{\pi}{2}$ as $r \rightarrow 0$ because of the presence of the singularity. Further, the general spin precession remains finite for all angles $0<\theta<\frac{\pi}{2}$.
\par
\begin{figure}
\centering
\begin{subfigure}{0.33\linewidth}
	\centering
	\includegraphics[width=\linewidth]{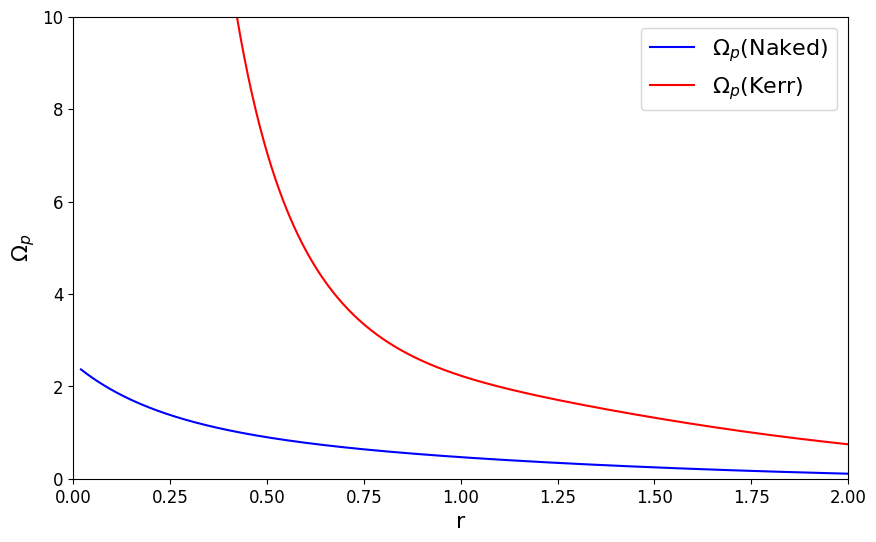}
	\caption{$\theta=\frac{\pi}{2},~k=0.2$}
\end{subfigure}%
\begin{subfigure}{0.33\linewidth}
	\centering
	\includegraphics[width=\linewidth]{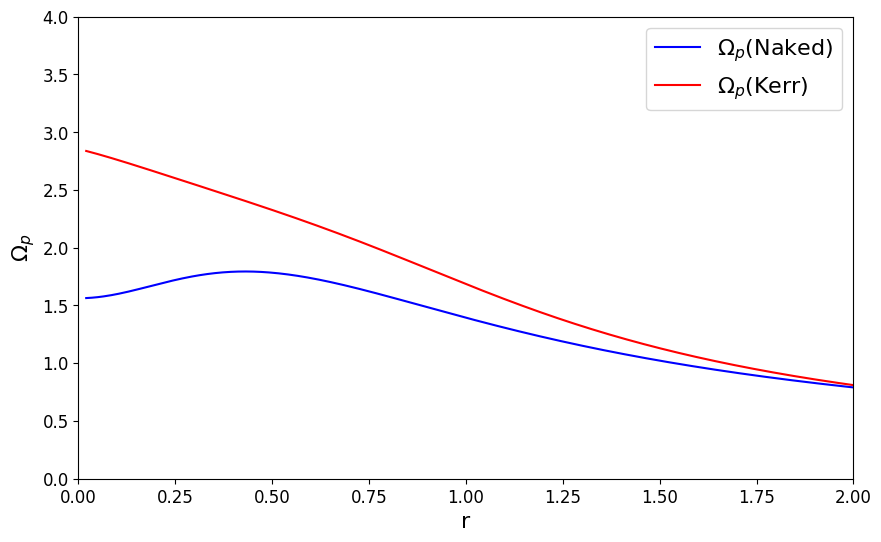}
	\caption{$\theta=\frac{\pi}{6},~k=0.2$}
\end{subfigure}%
\begin{subfigure}{0.33\linewidth}
	\centering
	\includegraphics[width=\linewidth]{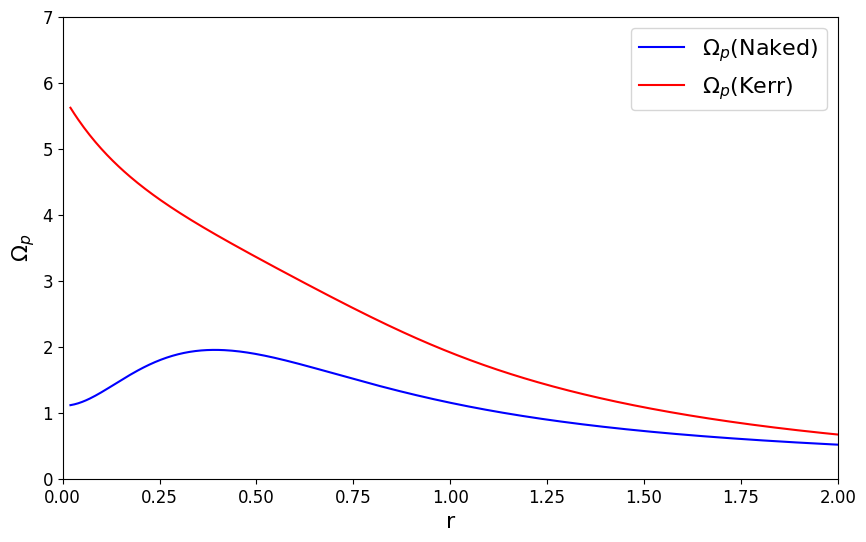}
	\caption{$\theta=\frac{\pi}{4},~k=0.2$}
\end{subfigure}
\begin{subfigure}{0.33\linewidth}
	\centering
	\includegraphics[width=\linewidth]{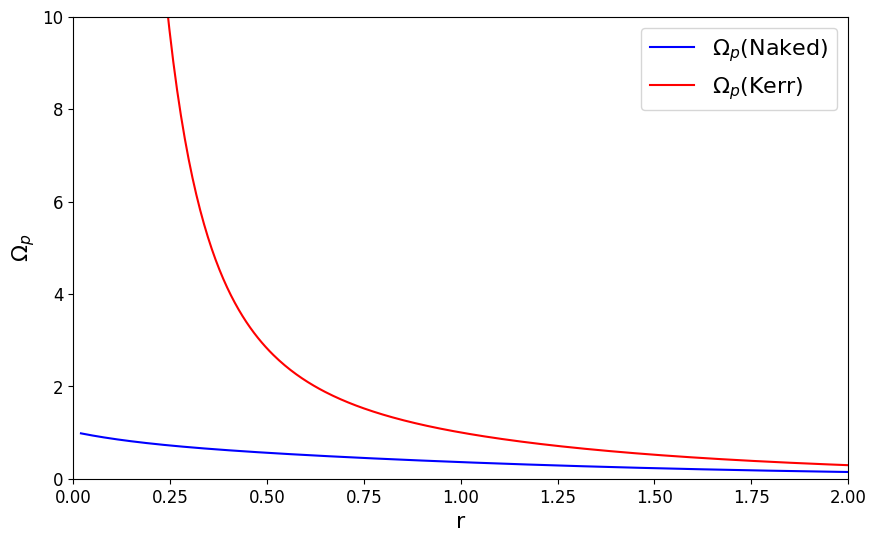}
	\caption{$\theta=\frac{\pi}{2},~k=0.5$}
\end{subfigure}%
\begin{subfigure}{0.33\linewidth}
	\centering
	\includegraphics[width=\linewidth]{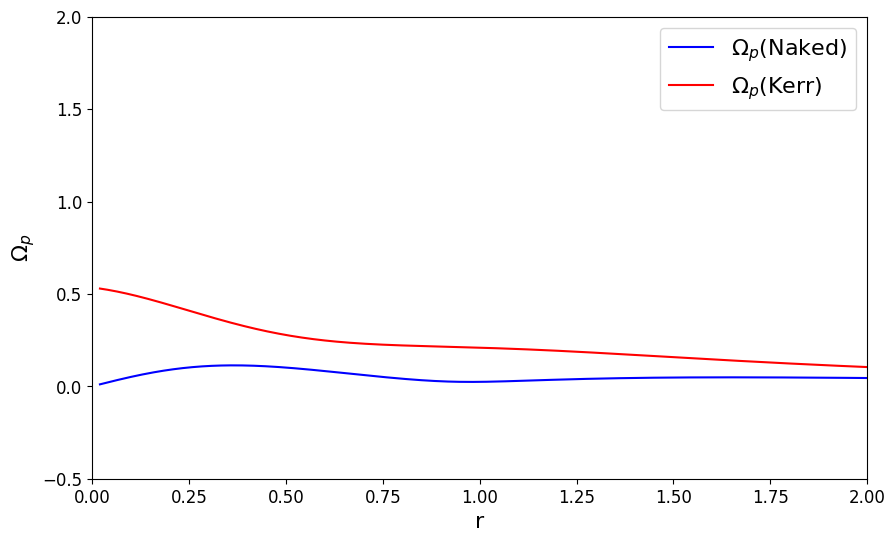}
	\caption{$\theta=\frac{\pi}{6},~k=0.5$}
\end{subfigure}%
\begin{subfigure}{0.33\linewidth}
	\centering
	\includegraphics[width=\linewidth]{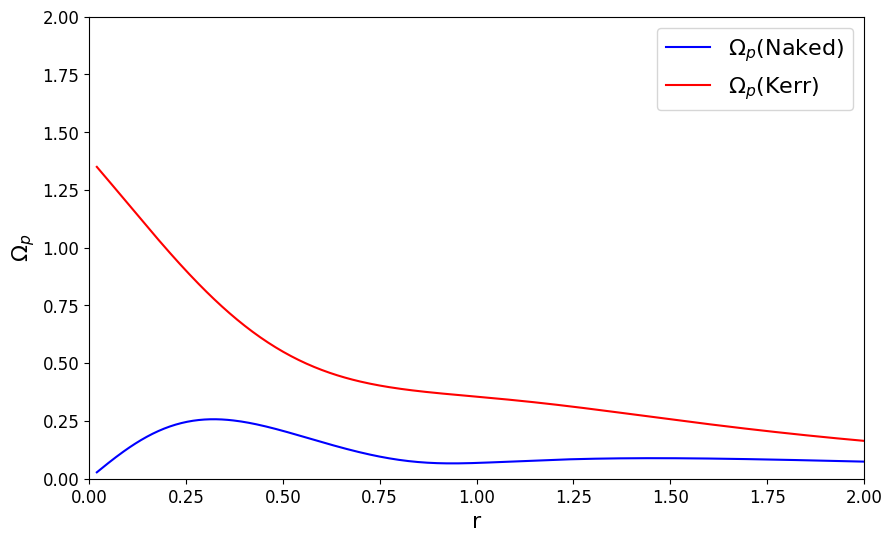}
	\caption{$\theta=\frac{\pi}{4},~k=0.5$}
\end{subfigure}
\caption{Magnitude of general spin precession frequency for Kerr naked singularity and RNS at $a=1.2$ and $M=1$. In the lower panel, the general spin precession for ZAMO is plotted. General spin precession for Kerr naked singularity diverges for the observer in the equatorial plane as $r \rightarrow 0$.}
\label{Fig.6}
\end{figure}

In Fig.~\ref{Fig.6}, we drew a comparison between the Kerr naked singularity described by (\ref{4.1}) with $a>M$ and RNS (\ref{2.1})  using the spin precession. To explain this, consider a gyroscope attached to an observer moving with an angular velocity. Suppose that the observer is moving toward the singularity along $\theta=\frac{\pi}{2}$. The central object will be the RNS if the spin precession remains finite as we approach the central object, otherwise, it is a Kerr naked singularity. For an observer along a direction other than $\theta=\frac{\pi}{2}$, we cannot differentiate between these two naked singularities.
\par
It is to be noted that we considered the larger value of spin parameter $a=1.2$ for RNS to make a comparison with the Kerr naked singularity. For small values of spin parameter ($a \leq 1$), we can easily distinguish it from a Kerr black hole. This is because the general spin precession in a Kerr black hole ($a \leq 1$) diverges at \textit{event horizon} while in the case of RNS, it remains finite except at the singularity as demonstrated in Fig.~\ref{Fig.1}.

\section{QPOs Frequencies in the Rotating Naked Singularity}
\label{section-4}
\begin{figure}
\centering
\begin{subfigure}{0.5\linewidth}
	\includegraphics[width=\linewidth]{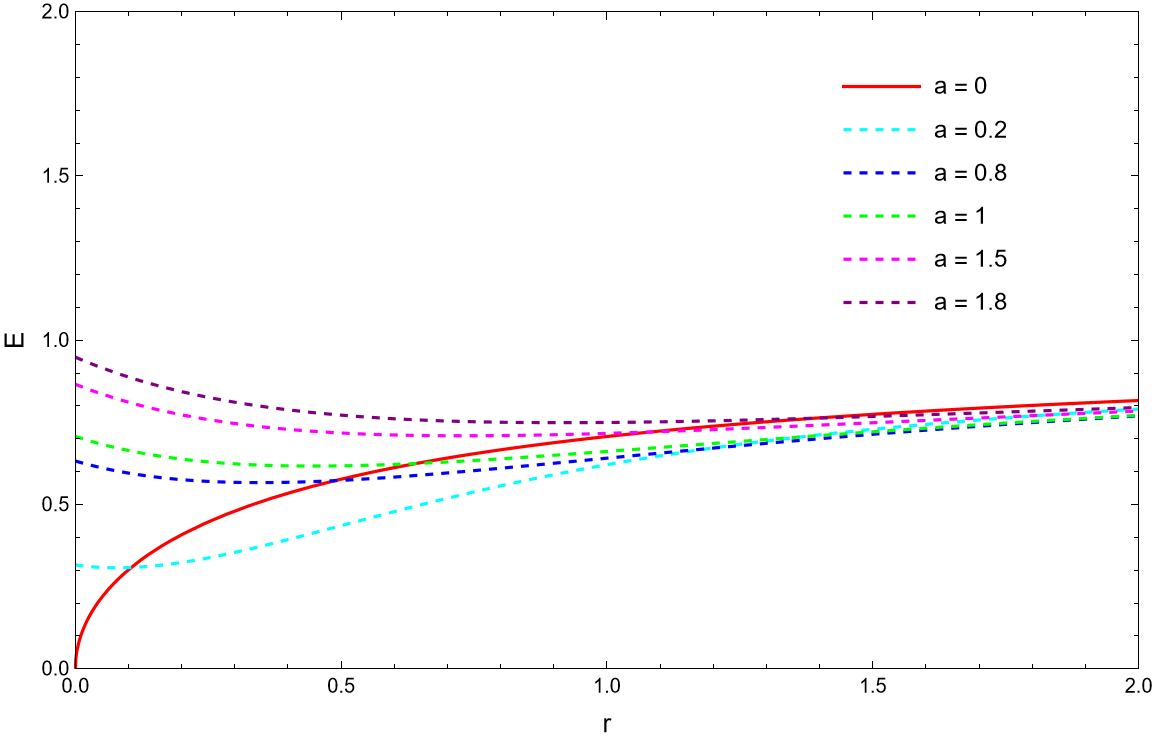}
	\caption{co-rotating}
\end{subfigure}%
\begin{subfigure}{0.5\linewidth}
	\includegraphics[width=\linewidth]{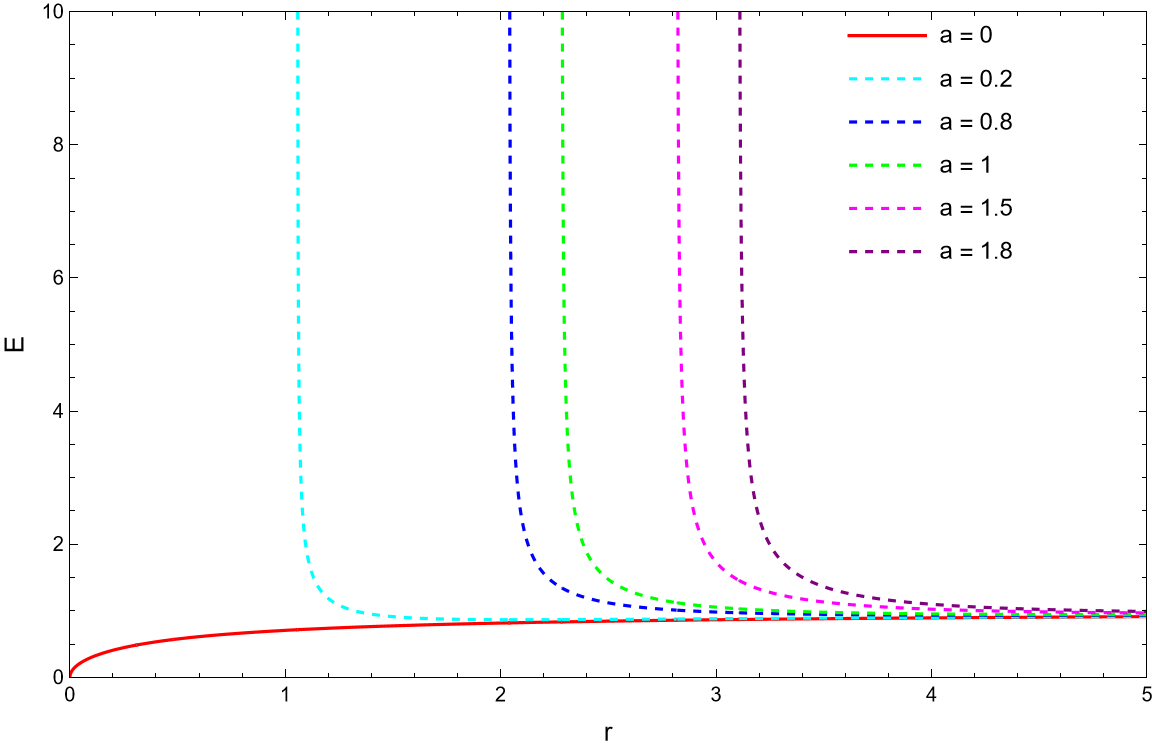}
	\caption{counter-rotating}
\end{subfigure}
\begin{subfigure}{0.5\linewidth}
	\includegraphics[width=\linewidth]{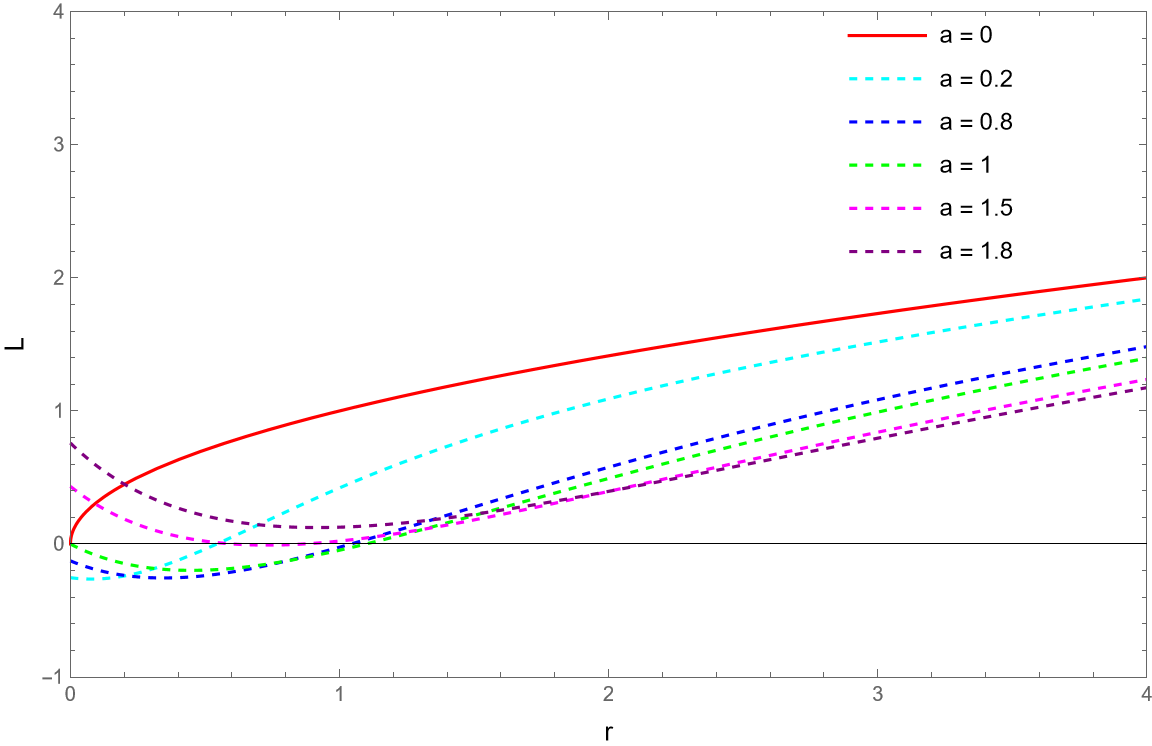}
	\caption{co-rotating}
\end{subfigure}%
\begin{subfigure}{0.5\linewidth}
	\centering
	\includegraphics[width=\linewidth]{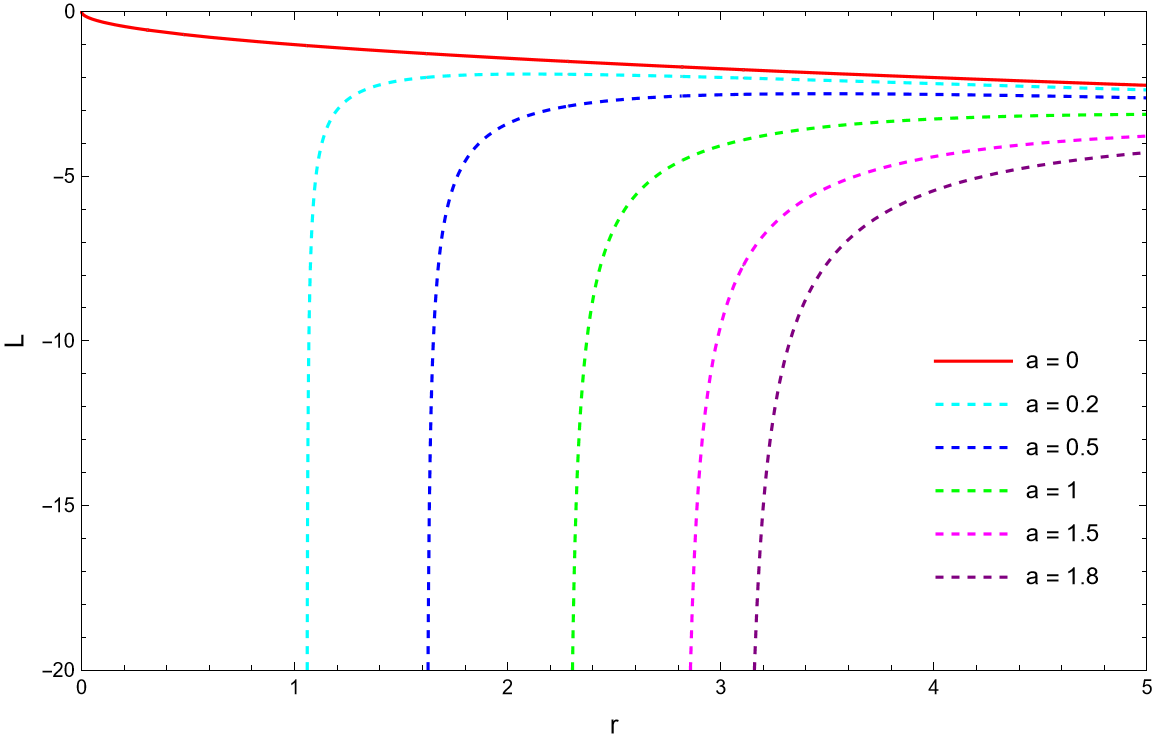}
	\caption{counter-rotating}
\end{subfigure}
\caption{Parameters $E$ and $L$ for different values of $a$ have been plotted for co-rotating and counter-rotating orbits taking $M=1$. For counter-rotating orbits, the curves for $E$ and $L$ diverge at a specific value of $r=r_\text{ph}$, known as the radius of the photon orbit.}
\label{Fig.7}
\end{figure}

\begin{figure}
\centering \includegraphics[width=0.5\linewidth]{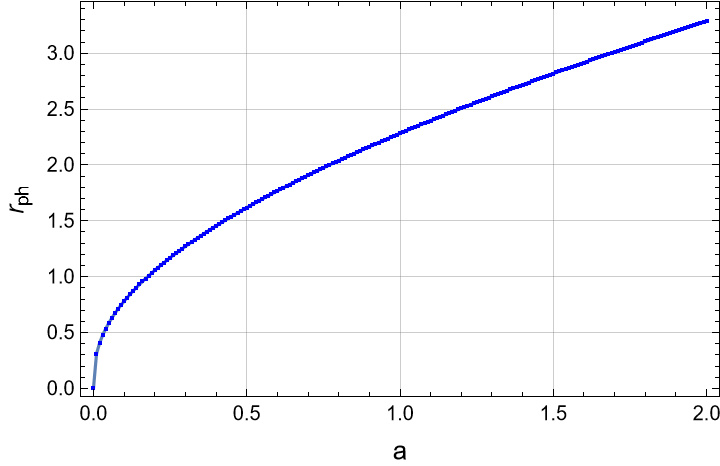}
\caption{Radius of photon orbit for counter-rotating orbit taking $M=1$.}
\label{Fig.photon}
\end{figure}

This section deals with deriving the governing equation for ISCO and checking the variation of the orbital, periastron precession, and nodal precession frequency functions for the RNS. A test particle is characterized by three fundamental frequencies in an equatorial circular orbit. These frequencies depend upon the metric of spacetime and generally we can write as
\begin{align}
\label{5.1}
\nu_{i}=\frac{\Omega_{i}}{2\pi},
\end{align}
with $i=\phi,r,\theta$.  Here $\nu_{\phi}$, $\nu_r$, and $\nu_\theta$ refer to Keplerian frequency (orbital frequency),  radial epicyclic frequency, and vertical epicyclic frequency, respectively. Radial (vertical) is the frequency of radial (vertical) oscillation around the mean orbit.
The stationary axi-symmetric metric is independent of coordinates $t$ and $\phi$, hence having two Killing isometries, hence two constants of motion can be defined, namely, specific energy $E$ and specific angular momentum $L$ given by
\begin{align}
\frac{p_{t}}{m}&=-E=\Dot{t}(g_{tt}+g_{t\phi}\Omega_{\phi}),
\end{align}
\begin{align}
\frac{p_{\phi}}{m}&=L=\Dot{t}(g_{t\phi}+g_{\phi\phi}\Omega_{\phi}),
\end{align}
where $\Omega_{\phi}=\Dot{\phi}/\Dot{t}=d\phi/dt$ is the orbital angular velocity. We assume a stationary axi-symmetric metric such that $g_{\beta\sigma,t}=0$ and $g_{\beta\sigma,\phi}=0$. For circular equatorial geodesic (having conditions $\dot{r} =0,\ \dot{\theta} =0,\ \ddot{r} = 0$), the geodesic equation $ \frac{d}{d\tau}(g_{\alpha\beta}\dot{x}^{\beta}) = \frac{1}{2}(\partial_{\alpha}g_{\beta\sigma})\dot{x}^{\beta}\dot{x}^{\sigma},$ becomes
\begin{align*}
g_{\phi\phi,r}\Omega^2_{\phi}+2g_{t\phi,r}\Omega_{\phi}+g_{tt,r}=0.
\end{align*}
From the above expression, we can write the orbital angular velocity using the quadratic formula as
\begin{align}
\Omega_{\phi}=\dfrac{-g_{t\phi,r}\pm\sqrt{g_{t\phi,r}^{2}-g_{tt,r}g_{\phi\phi,r}}}{g_{\phi\phi,r}}.
\end{align}
Orbital angular velocity, energy and angular momentum for RNS are respectively given by
\begin{align}
\Omega_{\phi }&=\pm\frac{\sqrt{M}}{\sqrt{(M+r)^3}\pm a \sqrt{M}},\label{4.15}\\
E&= \frac{\pm a\sqrt{M} (M+r)^2+r^2 \sqrt{(M+r)^3}}{(M+r)^2 \sqrt{\pm2 a \sqrt{M} \sqrt{(M+r)^3}+r^3}},\label{4.16}\\
L&=\pm \frac{\sqrt{M} \left[a^2 (M+r)^2\mp a \sqrt{M} \sqrt{(M+r)^3} (M+2 r)+r^2 (M+r)^2\right]}{(M+r)^2 \sqrt{\pm2 a \sqrt{M} \sqrt{(M+r)^3}+r^3}},\label{4.17}
\end{align}
where the upper and lower signs correspond to co-rotating and counter-rotating orbits, respectively. In Fig.~\ref{Fig.7}, plots of energy and angular momentum are given.
The curves of $E$ and $L$ for counter-rotating orbits diverge at a radius called the radius of photon orbit $r_\text{ph}$. From the vanishing denominators of (\ref{4.16}) and (\ref{4.17}), we get the equation of photon orbit with radius $r_\text{ph}$ given by,
\begin{align}
r_\text{ph}^3 -2 a \sqrt{M} \sqrt{(M+r_\text{ph})^3}=0.
\end{align}
It is a cubic equation and can be solved numerically to obtain the roots of $r$ for a specified value of $a$. In Fig.~\ref{Fig.photon}, we have plotted the numerical roots that are physically relevant. It was also demonstrated in \cite{patel_light_2022} that the photon sphere can only exist for counter-rotating orbits, unlike the Kerr spacetime.
\par
Using the normalization condition $g_{\alpha\beta}u^{\alpha}u^{\beta}=-1$, one can write
\begin{align}
V_\text{eff}(r,\theta)=g_{rr}\Dot{r}^{2}+g_{\theta\theta}\Dot{\theta}^{2},
\end{align}
where $r=r(\tau)$ and $\theta=\theta(\tau)$, while the effective potential $V_\text{eff}$ is given by
\begin{align}
\label{5.10}
V_\text{eff}(r,\theta)=\dfrac{E^{2}g_{\phi\phi}+2ELg_{t\phi}+L^{2}g_{tt}}{(g_{t\phi})^{2}-g_{tt}g_{\phi\phi}}-1.
\end{align}
For circular equatorial motion ($\theta=\frac{\pi}{2}$, and $r_{0}$ is constant) \cite{liu_testing_2023}
\begin{align}
\label{4.5}
U_\text{eff}(r)|_{r=r_{0}}=0, \frac{dU_\text{eff}(r)}{dr}|_{r=r_{0}}=0 \Rightarrow V_\text{eff}(r)|_{r=r_{0}}=0, \frac{dV_\text{eff}(r)}{dr}|_{r=r_{0}}=0,
\end{align}
where $U_\text{eff}(r)=V_\text{eff}(r)/g_{rr}$. If a test particle is perturbed from its equilibrium position, then it will start to oscillate around the mean position. If $\eta_{r}$ and $\eta_{\theta}$ are small displacements along the radial and angular directions about the mean orbit, i.e., $r=r_{0}+\eta_{r}$ and $\theta=\theta_{0}+\eta_{\theta}$, then after perturbation, we obtain the following second order ODEs (to linear order in $\eta$) \cite{Rayimbaev_2022,Boshkayev2023QuasiperiodicOF}
\begin{align}
\frac{{d^2\eta_{r}}}{dt^2}+\Omega_{r}^{2}\eta_{r}=0,
\quad
\frac{{d^{2}\eta_{\theta}}}{dt^{2}}+\Omega_{\theta}^{2}\eta_{\theta}=0,
\end{align}
where the square of frequencies for radial and angular coordinates are given by \cite{narzilloev2021dynamics}
\begin{align}
\label{5.14}
\Omega_{r}^{2}=\frac{-1}{2 g_{rr} \Dot{t}^{2}}\frac{\partial^{2} V_\text{eff}}{\partial r^{2}}|_{\theta=\frac{\pi}{2}},
\end{align}
\begin{align}
\label{5.15}
\Omega_{\theta}^{2}=\frac{-1}{2 g_{\theta\theta} \Dot{t}^{2}}\frac{\partial^2 V_\text{eff}}{\partial \theta^2}|_{\theta=\frac{\pi}{2}}.
\end{align}
For a stable circular equatorial orbit $d^{2}U_\text{eff}(r)/dr^2|_{r=r_{0}}<0$ implying  $d^{2}V_\text{eff}(r)/dr^2|_{r=r_{0}}<0$ i.e., the test particle is located at the maxima of the given term to ensure the radial stability\footnote{Considering the representation $\Dot{r}^{2} = U_{\text{eff}}$, a stable circular equatorial orbit have the conditions
$U_\text{eff}(r)\big|_{r=r_{0}}=0, \frac{dU_\text{eff}(r)}{dr}\big|_{r=r_{0}}=0, \frac{d^{2}U_\text{eff}(r)}{dr^{2}}\big|_{r=r_{0}}<0$ while the representation $\dot{r}^2 = E^2 - U_{\text{eff}}$ demands $U_\text{eff}(r)\big|_{r=r_{0}}=E^2,  \frac{d U_\text{eff}(r)}{d r}\big|_{r=r_{0}}=0, \frac{{d^{2}U_\text{eff}}(r)}{dr^{2}}\big|_{r=r_{0}}>0$ for a stable circular equatorial orbit as demonstrated in \cite{Azreg-Ainou:qpos,AzregAinou2019}.}. Plugging (\ref{5.10}) in (\ref{5.14}) and (\ref{5.15}) we can obtain $\Omega_{r}$ and $\Omega_{\theta}$ \cite{Wu:2023wld}. Hence, using (\ref{5.1}), we can write the fundamental frequencies of the test particle in the equatorial circular orbit as follows
\begin{eqnarray}
\nu_\phi&=&\pm\frac{1}{2\pi}\frac{\sqrt{M}}{\sqrt{(M+r)^3}\pm a \sqrt{M}}, \label{5.16} \\
\nu_{r}&=& \nu_{\phi}\left(\sqrt{\frac{r^{3}}{(M+r)^{3}}-\frac{3a^{2}}{r(M+r)}\pm\frac{8a\sqrt{M}}{\sqrt{(M+r)^{3}}}}\right), \\
\nu_{\theta}&=&\nu_{\phi}\left(\sqrt{1\mp\frac{2 a \sqrt{M}(M+2r)}{r^2 \sqrt{(M+r)}}+\frac{a^{2}(M^2+3 M r+3 r^2)}{r^{4}}}\right).\label{5.18}
\end{eqnarray}
The upper and lower signs correspond to co-rotating and counter-rotating orbits, respectively. It is to be noted that if the distant observer's measurements of the frequencies of small harmonic oscillations are given in physical units, then the corresponding dimensionless form must be extended by the factor $\frac{c^3}{GM}$\cite{Shahzadi:2021upd,Jusufi_2021}.  
\par
\begin{figure}
\centering
\begin{subfigure}{0.5\linewidth}
	\centering
	\includegraphics[width=\linewidth]{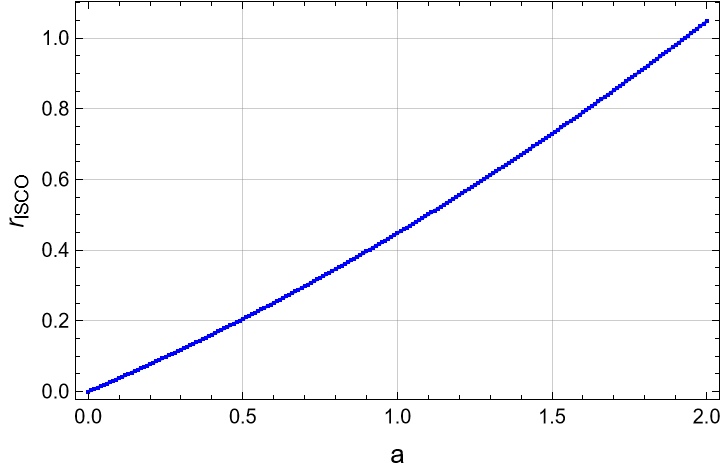}
	\caption{Co-rotating (prograde) orbits}
\end{subfigure}%
\begin{subfigure}{0.5\linewidth}
	\centering
	\includegraphics[width=\linewidth]{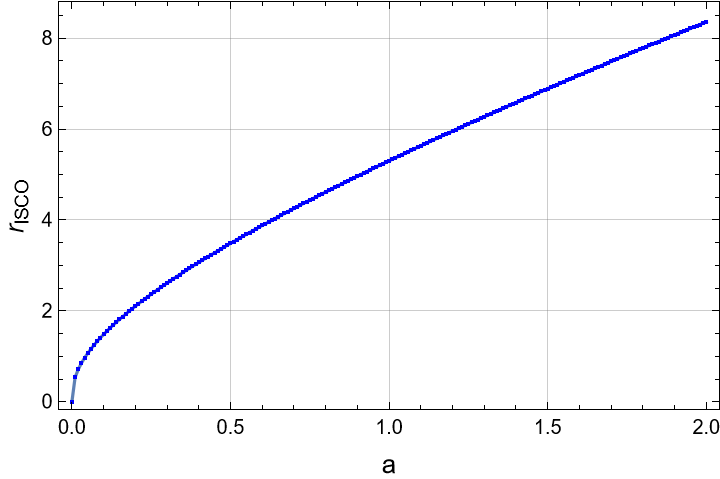}
	\caption{Counter-rotating (retrograde) orbits}
\end{subfigure}
\caption{The plots of ISCO radius versus $a$ are shown in the left and right panels for co-rotating and counter-rotating orbits, respectively, with $M=1$. It is evident that, for co-rotating orbits, the ISCO lies closer to the naked singularity compared to the counter-rotating orbits.
}
\label{Fig.8}
\end{figure}

The NNS case would recover by imposing $a=0$, i.e.,
\begin{align}
\nu_\phi&=\nu_{\theta}=\pm\frac{1}{2\pi}\frac{\sqrt{M}}{\sqrt{(M+r)^3}},\\
\nu_{r}&=\nu_{\phi}\sqrt{\frac{r^{3}}{(M+r)^{3}}}.
\end{align}
From the three fundamental frequencies, one can define the nodal precession frequency $\nu_{n}$ and periastron precession frequency $\nu_{p}$ as \cite{bambi_black_2017}
\begin{align}
\nu_{n}&=\nu_{\phi}-\nu_{\theta}, \label{nodal}\\
\nu_{p}&=\nu_{\phi}-\nu_{r}.\label{periastron}
\end{align}
Precession of orbit and orbital plane precession are measured by periastron precession frequency and nodal precession frequency, respectively. Orbital plane precession arises due to the rotation of spacetime and is also known as the Lense-Thirring precession frequency. Hence, nodal precession frequency is absent in NNS, being a non-rotating spacetime. Periastron precession is present in both NNS and RNS. It will be interesting to observe the behavior of the frequencies near the radius of ISCO. For ISCO, we have the condition  $\frac{{d^{2}U_{\text{eff}}(r)}}{dr^{2}}|_{r=r_{0}}=0,$ implying $\frac{{d^{2}V_{\text{eff}}(r)}}{dr^{2}}|_{r=r_{0}}=0,$ along with (\ref{4.5}). By using this condition, we obtain 
\begin{align*}
M r & \left(a^2 (M+r)^2+r^4\right)^2 \left[3 a^2 (M+r)^2\mp8 a \sqrt{M} r \sqrt{(M+r)^3}-r^4\right]=0.
\end{align*}
Since $a^2 (M+r)^2+r^4\geq 0$, hence the governing equation for ISCO becomes 
\begin{align}
\label{4.14}
r^4-3 a^2 (M+r)^2\pm8 a \sqrt{M} r \sqrt{(M+r)^3}=0,
\end{align}
where the upper and lower signs correspond to co-rotating and counter-rotating orbits, respectively. We can solve the above equation and obtain the numerical roots for a certain value of $a$. There are eight solutions, two of which are physically relevant. For the sake of simplicity, we will not provide the explicit expression, but the roots are plotted in Fig.~\ref{Fig.8}.
Note that, in \cite{Madan2022TidalFE}, the presence of radius of the ISCO was indicated for NNS and it was shown that [considering the representation $\dot{r}^2 = E^2 - U_{\text{eff}}(r)$] the stationary point of effective potential i.e., $\frac{L^2}{M}$ is the ISCO which extends up to the singularity at $r=0$ as $L$ approaches $0$. But it has been observed that $\frac{d^2U_\text{eff}}{dr^2}|_{r=\frac{L^2}{M}}=\frac{2 L^2 M^4 +2M^6}{(L^2+M^2)^2}>0$, which turns out to be the point of minima ensuring the stable orbit. Hence, no ISCO is present in NNS spacetime.
Also, the ISCO radius is located at the minimum of energy and angular momentum, i.e., the minimum of energy $E$ and the angular momentum $L$ are located at the same radius \cite{bambi_black_2017}. Hence, if we compute $\frac{dE}{dr}=0,$ and $\frac{dL}{dr}=0,$ the solution of these two equations is the same as the solution of (\ref{4.14}).
\par
\begin{figure}
\centering
\begin{subfigure}{0.5\linewidth}
	\includegraphics[width=\linewidth]{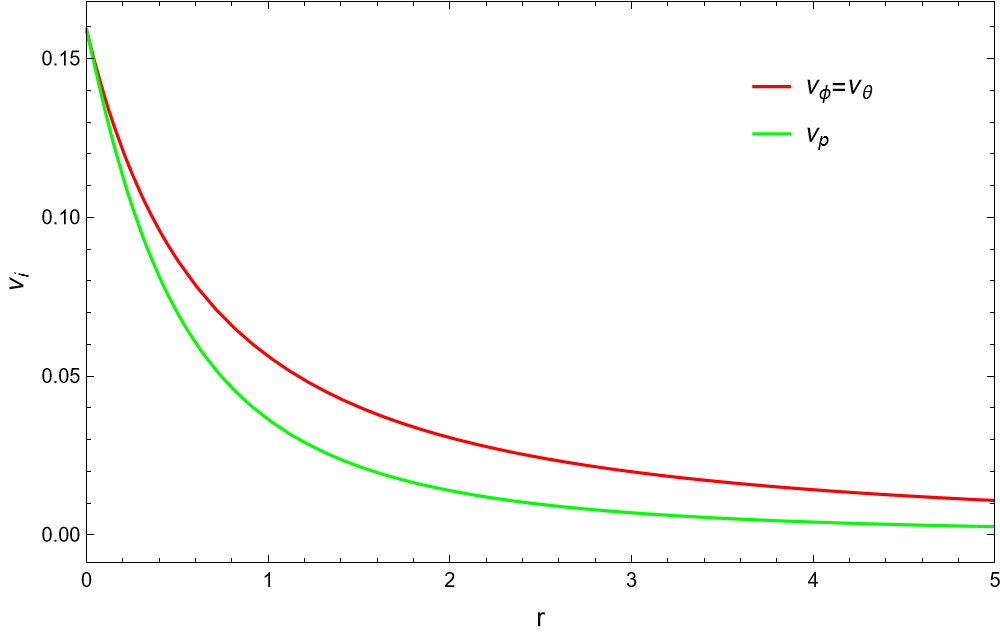}
	\caption{$a=0$}
\end{subfigure}%
\begin{subfigure}{0.5\linewidth}
	\includegraphics[width=\linewidth]{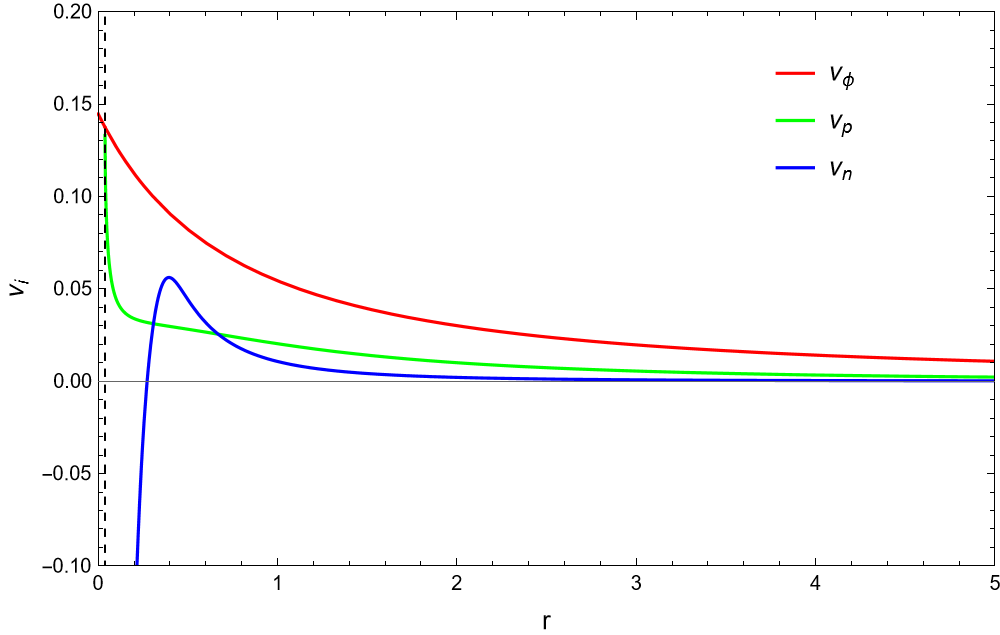}
	\caption{$a=0.1$}
\end{subfigure}
\begin{subfigure}{0.5\linewidth}
	\includegraphics[width=\linewidth]{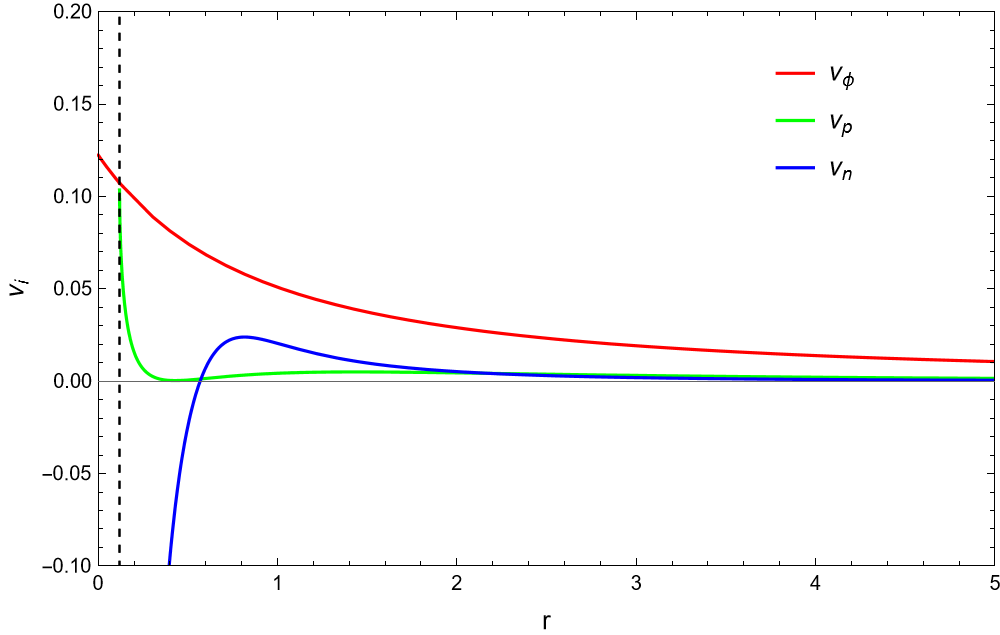}
	\caption{$a=0.3$}
\end{subfigure}%
\begin{subfigure}{0.5\linewidth}
	\includegraphics[width=\linewidth]{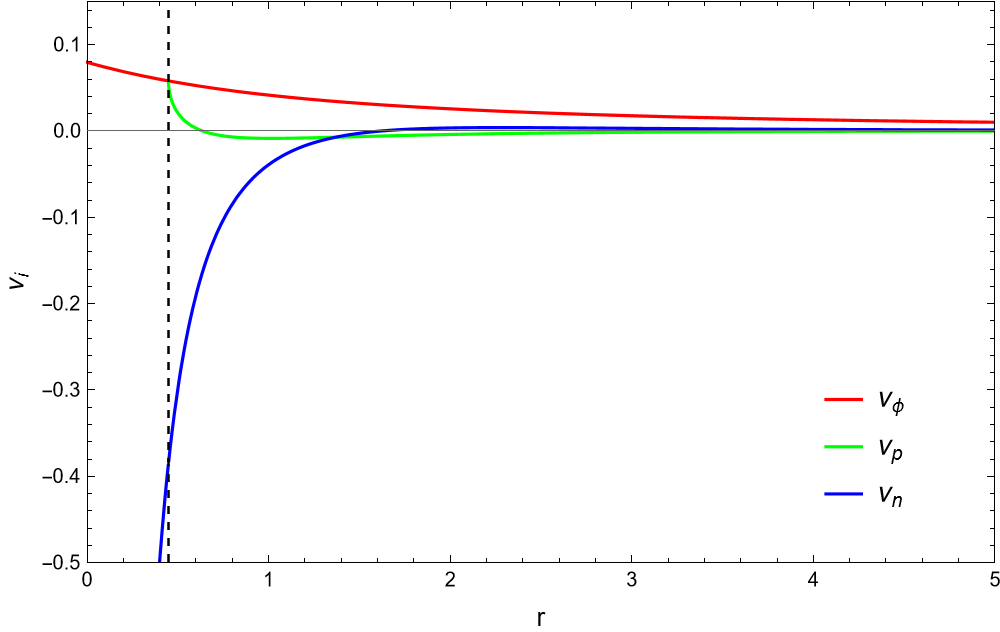}
	\caption{$a=1.0$}
\end{subfigure}
\caption{Variation of $\nu_{\phi}$, $\nu_{p}$, and $\nu_{n}$ for co-rotating orbits at $M=1$. The dotted lines represent the ISCO position. For NNS, orbital frequency and vertical epicyclic frequency become equal as shown in panel (a), which is why there is no nodal precession in NNS. For RNS, it can be observed that periastron precession vanishes at $r_{\text{ISCO}}$. $\nu_{n}$ increases, attains a certain peak, and then starts decreasing. The peak decreases with an increase in  $a$. }
\label{Fig.9}
\end{figure}
\begin{figure}
\centering
\begin{subfigure}{0.5\linewidth}
	\includegraphics[width=\linewidth]{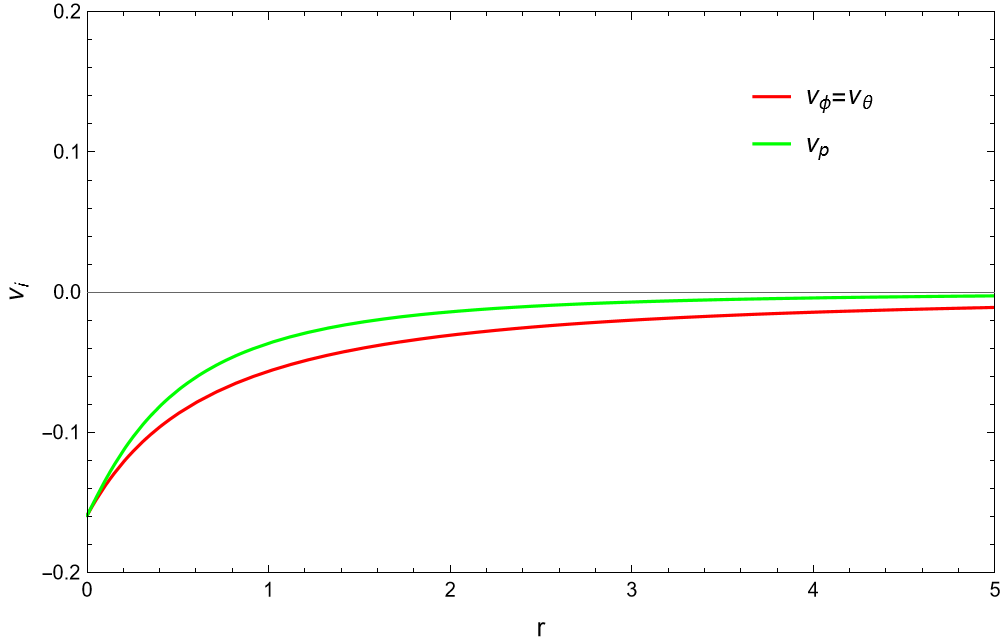}
	\caption{$a=0$}
	\label{Fig.10(a)}
\end{subfigure}%
\begin{subfigure}{0.5\linewidth}
	\includegraphics[width=\linewidth]{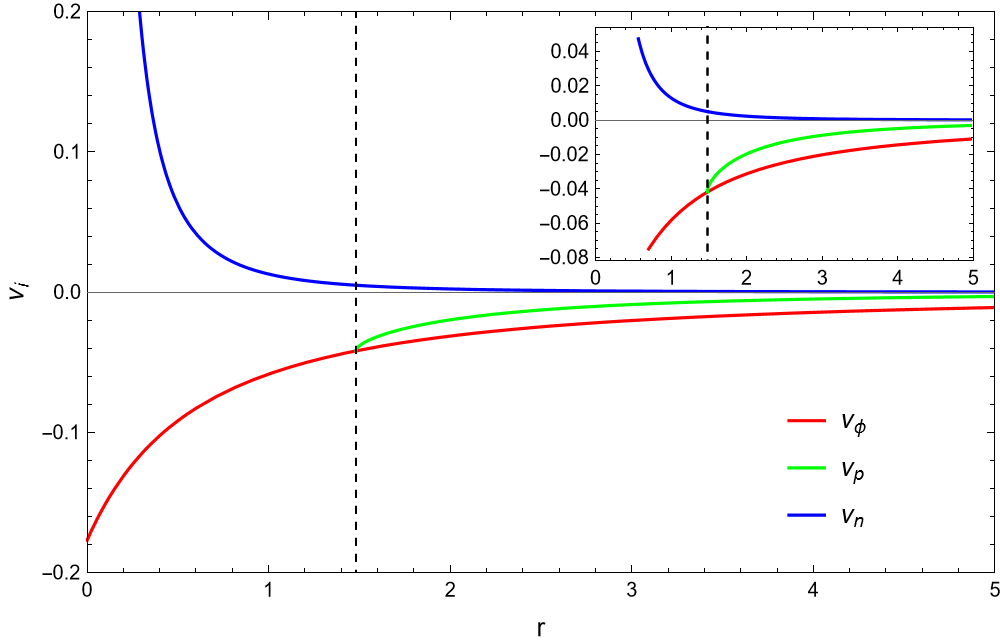}
	\caption{$a=0.1$}
	\label{Fig.10(b)}
\end{subfigure}
\begin{subfigure}{0.5\linewidth}
	\includegraphics[width=\linewidth]{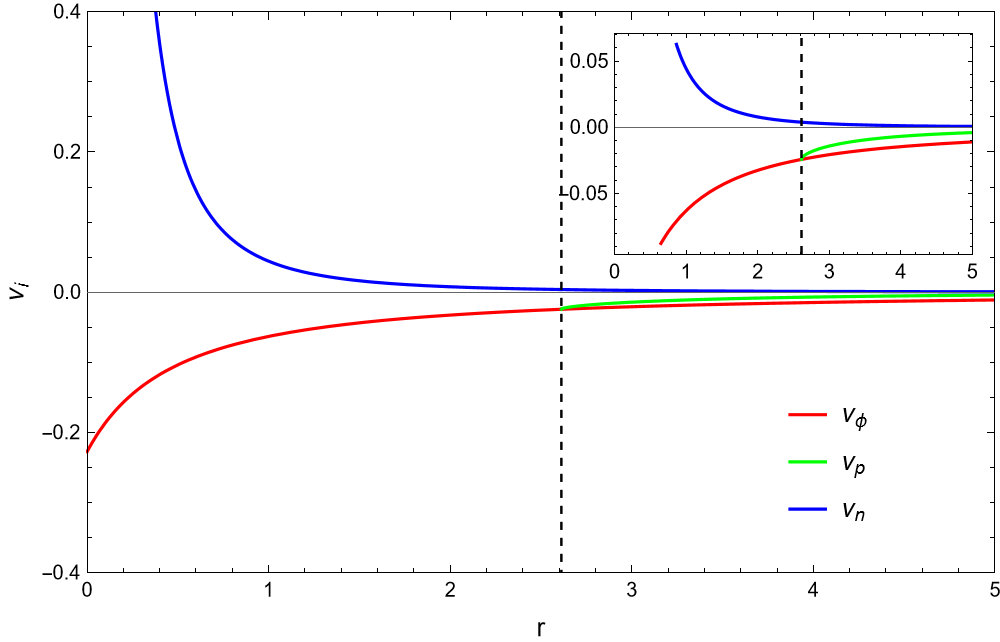}
	\caption{$a=0.3$}
	\label{Fig.10(c)}
\end{subfigure}%
\begin{subfigure}{0.5\linewidth}
	\includegraphics[width=\linewidth]{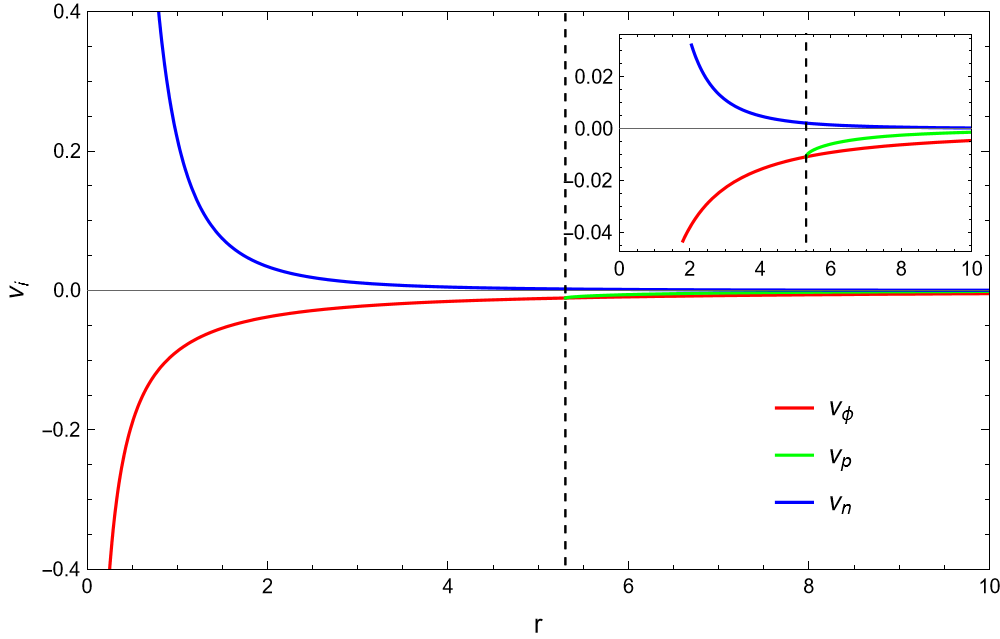}
	\caption{$a=1.0$}
	\label{Fig.10(d)}
\end{subfigure}
\caption{Variation of $\nu_{\phi}$, $\nu_{p}$, and $\nu_{n}$ for counter-rotating orbits at $M=1$. The dotted lines represent the ISCO
	position. In panel (a), $\nu_{\phi}=\nu_{\theta}$ and $\nu_{n}$ is absent for NNS. We have zoomed in the plots for RNS from (b) to (d), showing $\nu_{p}$ vanishes at $r_{\text{ISCO}}$. Unlike the co-rotating orbit, we do not observe a peak in $\nu_{p}$ for the counter-rotating orbit. Also, $\nu_{n}$ starts decreasing as we increase $a$.}
\label{Fig.10}
\end{figure}

To ensure that $E$ and $L$ are real numbers for a counter-rotating orbit, we impose the following constraint 
\begin{align}
r^{3}>2a\sqrt{M}\sqrt{(M+r)^3}>0.
\end{align}
It is to be noted that $\Omega_{r}^2(r_{ISCO})=0$. Hence $\Omega_{r}^{2}>0$ ensures the radial stability against small oscillations. Thus, the motion will be radially stable if
\begin{align}
r^4 -3 a^2 (M+r)^2\pm8 a \sqrt{M} r \sqrt{(M+r)^3}>0.
\end{align} 
The variation of orbital frequency, nodal precession frequency $v_{n}$, and periastron precession frequency $\nu_p$ are plotted in Fig.~\ref{Fig.9} and Fig.~\ref{Fig.10} for co-rotating and counter-rotating orbits, respectively. We can see that  $\nu_{p}$ vanishes at the radius of ISCO. As we approach the central object,  $\nu_{n}$ attains a certain peak at $r>r_{\text{ISCO}}$, and this peak decreases as $a$ increases. At a certain radius, $\nu_{n}$ vanishes and then becomes negative, indicating the reversion in the precession direction \cite{Rizwan_2019}. In Fig.~\ref{Fig.10}, we observe that the orbital frequency $\nu_{\phi}$ diverges in the case of $a = 1$, as the denominator of (\ref{5.16}) vanishes at $r=0$. Further, $\nu_{p}$ vanishes at $r_{\text{\text{ISCO}}}$ similar to the case of co-rotating orbit. However, it starts decreasing as $a$ increases. Note that $\nu_{n}$ does not show the reversion in the precession direction in this case. It decreases as $r$ increases and diverges as $r$ approaches $0$.

\section{Constraints on Parameters of Rotating Naked Singularity within Relativistic Precession Model}
\label{section-5}
In this section, we will utilize the RPM along with observed QPOs frequencies from five XRBs to constrain the parameter in RNS with the aid of the MCMC simulation method.
In RPM the upper HFQPO $\nu_{U}$, lower HFQPO $\nu_{L}$ and type-C LFQPO $\nu_{C}$ would correspond to the orbital frequency $\nu_{\phi}$, periastron precession frequency $\nu_p$, and nodal precession frequency $\nu_{n}$ respectively \cite{Stella:Proposed-RPM}. The set of three QPOs is given by
\begin{align}
\nu_{U} = \nu_{\phi},\quad \nu_{C} = \nu_{n},  \quad \nu_{L} = \nu_{p} .
\end{align}
It can be seen from (\ref{5.16}),(\ref{nodal}), and (\ref{periastron}) that the above frequencies expressing the RPM depend only on the mass, spin, and radius at which QPOs are produced. The MCMC algorithm is used within the Bayesian framework to estimate posterior distributions. Let \(\Theta\) represent the parameters, \(D\) represent observed data, and \(\mathscr{M}\) represent the statistical model. Then, the posterior distribution can be expressed as 
\begin{align}
\mathscr{P}(\Theta|D, \mathscr{M}) \propto \mathscr{P}(D|\Theta, \mathscr{M}) \Pi(\Theta|\mathscr{M}),
\end{align}
where, $\propto$ is the sign of proportionality, $\mathscr{P}(\Theta|D, M)$, $\Pi(\Theta|\mathscr{M})$ $\mathscr{P}(D|\Theta, \mathscr{M})$ are posterior, prior and likelihood respectively. Once the posterior probability of the parameter is determined, we can use the posterior mean, median, and mode to make inferences about the parameters. It is often advantageous to work with the logarithm of the posterior probability to avoid numerical underflow. We assume the priors to be Gaussian priors given as 
\begin{align}
\Pi(\theta_{i})= \frac{1}{\sigma \sqrt{2\pi}} e^{-\frac{1}{2}\left(\frac{\theta_{i}-\mu}{\sigma}\right)^2},
\end{align}
where $\theta_{i}$ are the parameters, $\mu$ and $\sigma$ denote mean and standard deviation of the distribution respectively. A rich literature is available on the mechanism of MCMC \cite{montehogg2018data,sharma2017markov} and its implementation within RPM \cite{masoumeh,Shaymatov_2023}. Having set the basics, we can now perform MCMC to obtain the constraints by implementing \textit{emcee} \cite{emcee}. 
\begin{table}
\centering
\renewcommand{\arraystretch}{1.5} 
\setlength{\tabcolsep}{6pt} 
\begin{tabular}{c|c|c|c|c|c}
	\hline
	\hline
	& GRO J1655-40 & XTE J1550-564 & XTE J1859+226 & GRS 1915+105 &  H1743-322 \\
	\hline
	$\nu_\phi$ (Hz) & $441\pm2$ \cite{Bambi:testing-rpm} & $276\pm3$ \cite{Azreg-Ainou:qpos} & $227.5^{+2.1}_{-2.4}$\cite{Motta:XTE226} & $168 \pm 3$ \cite{Azreg-Ainou:qpos} & $240 \pm 3$ \cite{Ingram:SolutionsRPM} \\
	$\nu_{\text{per}}$ (Hz) & $298\pm4$ \cite{Bambi:testing-rpm} & $184\pm5$ \cite{Azreg-Ainou:qpos} & $128.6^{+1.6}_{-1.8}$ \cite{Motta:XTE226} & $113 \pm 5$ \cite{Azreg-Ainou:qpos} & $156^{+9}_{-5}$ \cite{Ingram:SolutionsRPM} \\
	$\nu_{\text{nod}}$ (Hz) & $17.3\pm0.1$ \cite{Bambi:testing-rpm} & - & $3.65\pm0.01$\cite{Motta:XTE226} & - & $9.44 \pm 0.02$ \cite{Ingram:SolutionsRPM}\\
	\hline
	\hline
\end{tabular}
\caption{The observed frequencies of QPOs from the X-ray binaries selected for analysis.}
\label{tab:1}
\end{table}

\begin{table}
\centering
\renewcommand{\arraystretch}{1.5} 
\setlength{\tabcolsep}{6pt} 
\begin{tabular}{l|cc|cc|cc|cc|cc}
	\hline
	\hline
	\toprule
	Parameters & \multicolumn{2}{c|}{GRO J1655-40} & \multicolumn{2}{c|}{XTE J1550-564} & \multicolumn{2}{c|}{XTE J1859+226} & \multicolumn{2}{c|}{GRS 1915+105} & \multicolumn{2}{c}{H1743-322} \\
	& $\mu$ & $\sigma$ & $\mu$ & $\sigma$ & $\mu$ & $\sigma$ & $\mu$ & $\sigma$ & $\mu$ & $\sigma$ \\
	\midrule
	$M(M_{\odot})$ & 5.307 & 0.066 & 9.10 & 0.61 & 7.85 & 0.46 & 12.41 & 0.62 & 9.29 & 0.46 \\
	$a/M$ & 0.286 & 0.003 & 0.34 & 0.007 & 0.149 & 0.005 & 0.29 & 0.015 & 0.27 & 0.013 \\
	$r/M$ & 5.677 & 0.035 & 5.47 & 0.12 & 6.85 & 0.18 & 6.10 & 0.30 & 5.55 & 0.27 \\
	\bottomrule
	\hline
	\hline
\end{tabular}
\caption{$\mu$ and $\sigma$ values selected for Gaussian priors \cite{liu2023constraints}.}
\label{Tab.Priors}
\end{table}
\begin{figure}
\centering
\includegraphics[scale=0.45]{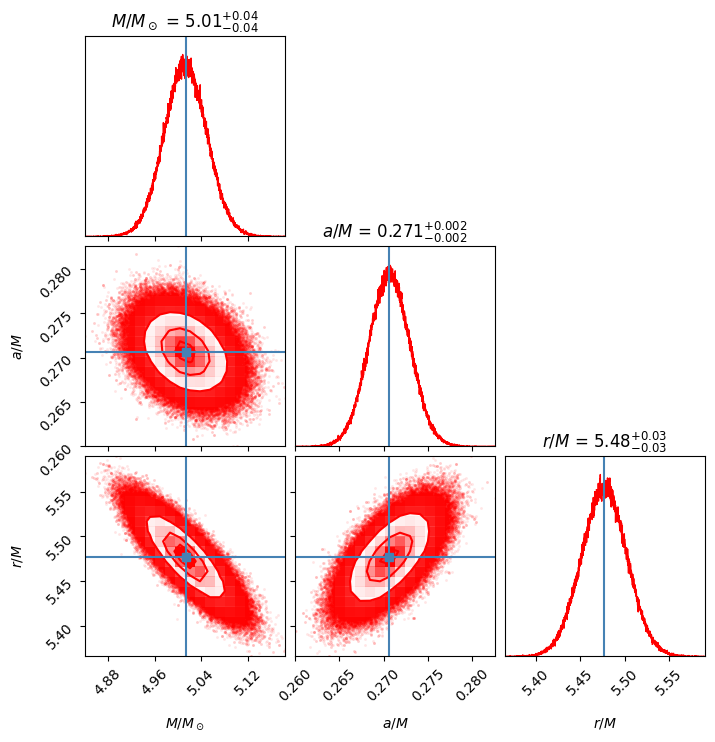}
\caption{One-dimensional posterior distribution for each parameter along diagonal and contour plots showing the correlation on other panels for the parameters obtained from the MCMC simulation for GRO J1655-40.}
\label{fig:constraints-Gro}
\end{figure}
\begin{figure}
\centering
\begin{subfigure}{0.5\linewidth}
	\centering
	\includegraphics[width=\linewidth]{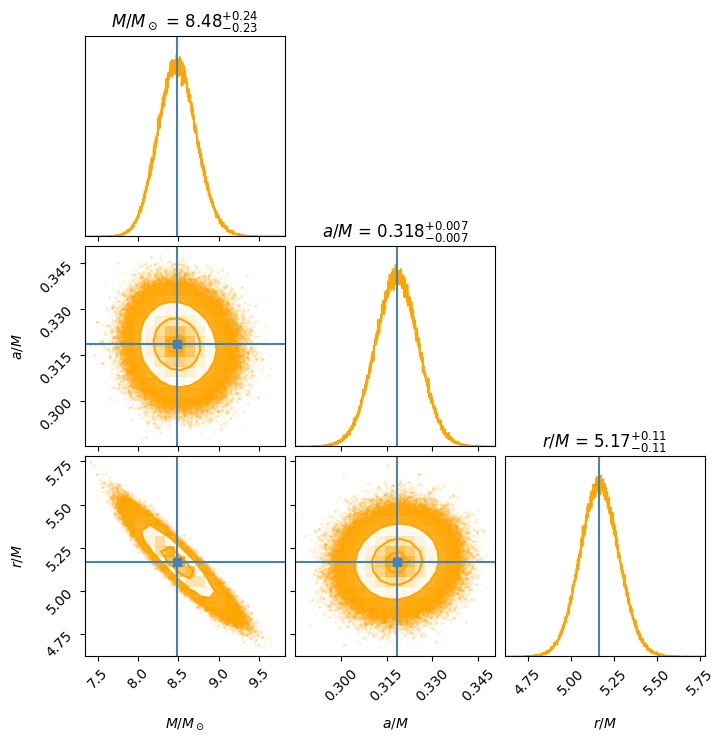}
	\caption{}
\end{subfigure}%
\begin{subfigure}{0.5\linewidth}
	\centering
	\includegraphics[width=\linewidth]{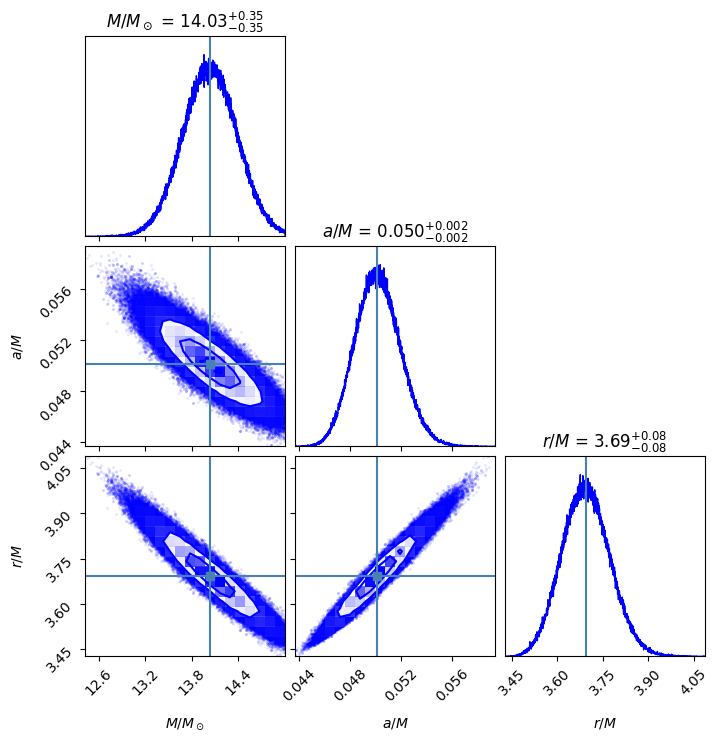}
	\caption{}
\end{subfigure}  
\begin{subfigure}{0.5\linewidth}
	\centering
	\includegraphics[width=\linewidth]{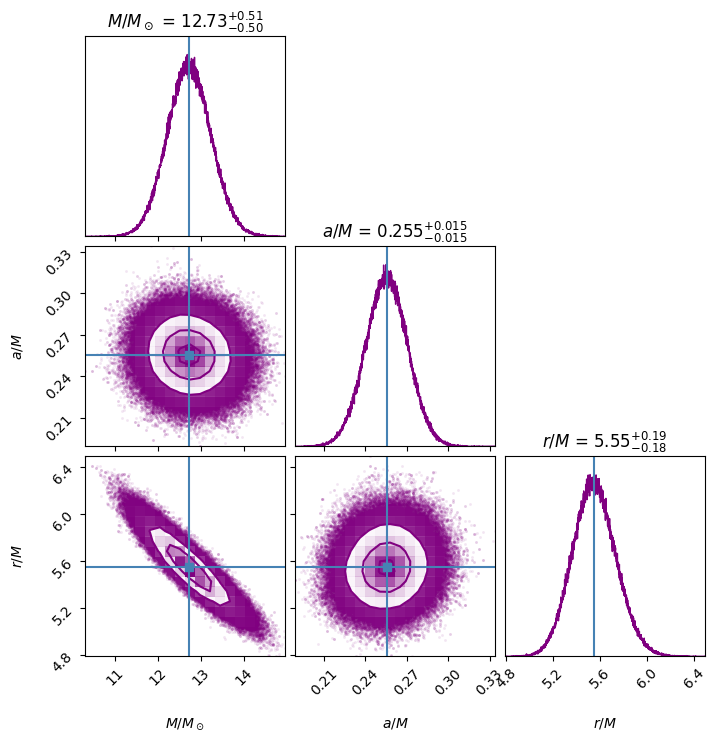}
	\caption{}
\end{subfigure}%
\begin{subfigure}{0.5\linewidth}
	\centering
	\includegraphics[width=\linewidth]{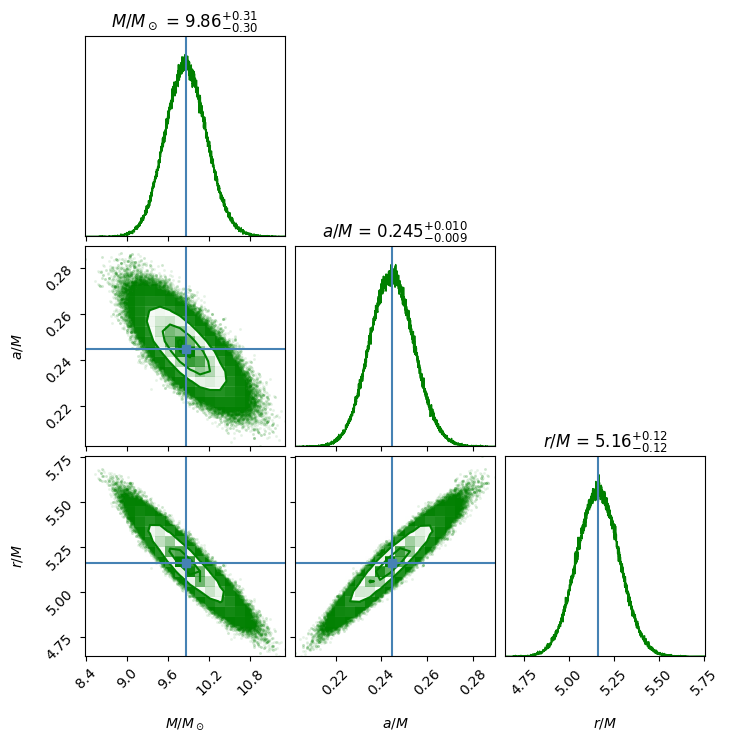}
	\caption{}
\end{subfigure}
\caption{Posterior distributions of the MCMC simulation for the parameters of RNS with
	XTE J1550-564 (orange contours), XTE J1859+226 (blue contours), GRS 1915+105 (purple contours), and H1743-322 (green contours). The one-dimensional posterior distributions for each parameter are shown on the diagonal, while the off-diagonal contour plots represent the two-dimensional correlations. The blue lines represent the mean.}
\label{fig:qpos}
\end{figure}
\begin{table}
\centering
\renewcommand{\arraystretch}{1.5} 
\setlength{\tabcolsep}{6pt} 
\begin{tabular}{c|c|c|c|c|c}
	\hline
	\hline
	& GRO J1655-40 & XTE J1550-564 & XTE J1859+226 & GRS 1915+105 &  H1743-322 \\
	\hline
	$M/M_\odot$ & $5.01 \pm 0.04$ &$8.48^{+0.24}_{-0.23}$& $14.03^{+0.35}_{-0.35}$ & $12.73^{+0.51}_{-0.50}$ & $9.86^{+0.31}_{-0.30}$  \\
	$a/M$ & $0.271 \pm 0.002$  & $0.318 \pm 0.007$ & $0.050 \pm 0.002$ & $0.255 \pm 0.015$ & $0.245^{+0.010}_{-0.009}$  \\
	$r/M$ & $5.48 \pm 0.03$& $5.17 \pm 0.11$ & $3.69 \pm 0.08$ & $5.55^{+0.19}_{-0.18}$ & $5.16 \pm 0.12 $\\
	$r_\text{ISCO}/M$ & $0.1069\pm0.0008$ & $0.1265 \pm 0.0029$  & $0.0189 \pm 0.07$  & $0.1003 \pm 0.0062$ & $0.0962^{+0.0041}_{-0.0004}$  \\
	\hline
	\hline
\end{tabular}
\caption{Parameter estimates as measured using MCMC within RPM. $r/M$ is the radius at which QPOs are emitted. $r_\text{ISCO}$ is calculated for the corresponding spin parameter $a$.}
\label{tab:2}
\end{table}
Observed frequencies along with their standard deviation for five binary systems are given in Table.~\ref{tab:1}. We set $\mu$ and $\sigma$ corresponding to each parameter for Gaussian priors given in Table.~\ref{Tab.Priors} taken from \cite{liu2023constraints,Liu2023}. We performed MCMC sampling with $10^5$ iterations to explore the posterior distribution of the model parameters. Finally, we obtain the best outcomes to examine a practical physical parameter region with the aid of MCMC simulation methods for RNS as shown in Fig.~\ref{fig:constraints-Gro} and Fig.~\ref{fig:qpos}. The shaded regions in contour plots show the $ 68 \% $, $90 \%$, and $95 \% $ confidence levels of the posterior probability density distributions. These figures show the constraints on the parameters of RNS from the current observations of QPOs within the relativistic precession model. We have summarized the RNS parameters and associated quantities in Table.~\ref{tab:2}. The analysis highlights a negative correlation between mass and radius for QPOs from all XRBs. A positive correlation exists between the spin parameter and radius in the case of GRO J1655-40, H1743-322, and XTE J1859+226. The mass and spin parameters are negatively correlated in H1743-322 and XTE J1859+226. 
\par
The spin parameter estimate for GRO J1655-40 is consistent with the one inferred in \cite{Bambi:testing-rpm} in which frequencies were obtained by considering the Kerr spacetime. The measurement of the mass reported in \cite{Bambi:testing-rpm} is in agreement with \cite{quiescent-meas} but has a smaller uncertainty, and the spin parameter, though obtained with high precision using RPM, is inconsistent with the value found in \cite{Shafee:2005ef}. Other estimates for spin and mass parameters are reported in \cite{Motta:precise-mass} for GRO J1655-40. The emission radius at which three QPOs are produced in RNS is found to be $5.48 \pm 0.03$, which is closer to the one found in \cite{Motta:precise-mass} given by $5.68 \pm 0.03$. The emission radius is also closer to the central object for RNS for XTE J1859+226 and XTE J1550-564 compared to the emission radius for the Kerr spacetime as obtained in \cite{Motta:XTE226,Motta:XTE564}. Spin parameter estimates for H1743-322 and XTE J1550-564 are consistent with \cite{H17} and \cite{continum564,Motta:XTE564} respectively. While the mass and spin parameter estimates derived from our MCMC simulations within the RNS spacetime framework show variation from those found in the prior studies using the Kerr spacetime, the differences are relatively minor for all XRBs (except  XTE J1859+226). This suggests that while the choice of spacetime model does influence the parameter estimates, the effect is not drastic. There is a significant deviation in estimates of XTE J1859+226, demonstrating a substantial difference when comparing the naked singularity model to the Kerr black hole model \cite{Motta:XTE226}. 
\begin{table}
\centering

\begin{tabular}{l |l| c}
	\toprule
	\hline
	Source & Method & a/M \\
	\midrule
	GRO J1655-40 & RPM & $0.290 \pm 0.003$ \cite{Motta:precise-mass}\\
	GRO J1655-40 & RPM & $0.286 \pm 0.006$\cite{Bambi:testing-rpm} \\
	XTE J1859+226 & RPM & $0.149 \pm 0.005$ \cite{Motta:XTE226}\\
	XTE J1550-564 & RPM & $0.34 \pm 0.01$ \cite{Motta:XTE564} \\
	H1743–322 & RPM & $\geq 0.21$ \cite{Ingram:SolutionsRPM} \\
	GRO J1655-40 & Continuum fitting & $0.7 \pm 0.1$\cite{Shafee:2005ef} \\
	GRO J1655-40 & Fits to the spectra   & $0.98$ \cite{gro-reflection} \\
	XTE J1550-564 & Continuum fitting & $0.34$\cite{continum564} \\
	XTE J1859+226 & X-ray reflection spectroscopy  & $0.986^{+0.001}_{-0.004}$ (68\% CL) \cite{XTE226spinSpectro}\\
	H1743–322 & Fits to RXTE spectra & $0.2 \pm 0.3$ (68\% CL) \cite{H17} \\
	GRS 1915+105 & Spectral analysis  & $> 0.98$ \cite{GRSspectralAnalys}\\
	
	\bottomrule
	\hline
\end{tabular}
\caption{Dimensionless spin parameters ($a/M$) for various sources given in literature. The first five rows correspond to the parameter estimated within RPM by considering the central object to be a Kerr black hole. }
\label{Tab.4}
\end{table}

\section{Conclusion and Discussion}
\label{section-6}

Studies show that inhomogeneous dust collapse will lead to singularity formation without a horizon and hence be visible to us \cite{homo2,homo3,homo4}. The physical properties of a naked singularity differ from a black hole. Thus, we investigated the general spin precession frequency of a test gyroscope in RNS spacetime. General spin precession in RNS remains finite everywhere for all observers except at the singularity itself. This behavior makes RNS different from the Kerr naked singularity and helps us distinguish between them. In weak field approximation, the frame-dragging effect induced by RNS is equivalent to the frame-dragging effect of Kerr spacetime as $\Vec{\Omega}_{\text{LT}}^{(\text{RNS})}(\text{weak})=\Vec{\Omega}_{\text{LT}}^{(\text{Kerr})}(\text{weak})$. Hence RNS can be thought of as a Kerr mimicker. We can conclude that in a weak field limit of a rotating compact object, an astronaut cannot differentiate between RNS and Kerr spacetime, as his/her gyro spin measurement would show the same precession in both cases. Asymptotically, the NNS resembles Schwarzschild spacetime as $\Delta\phi_{\text{geodetic}, \odot}^{(\text{NNS})}=\Delta\phi_{\text{geodetic}, \odot}^{(\text{Sch})}$. Hence, NNS can be thought of as a Schwarzschild mimicker.
\par
While this study focuses on spin precession for two specific classes of observers, i.e., stationary observers (who follow circular orbits) and static observers, we acknowledge that astrophysical objects, such as pulsars, often exhibit more complex dynamical behaviors. For example, as highlighted in \cite{astro}, pulsars in equatorial circular orbits around Kerr black holes experience spin precession effects that can substantially alter their observed pulse frequencies, particularly as they approach the event horizon. This could make it possible to identify horizons, ergoregions, or even test the no-hair theorem for black holes based on deviations in pulsar spin precession measurements. Our present study establishes how naked singularities mimic black holes asymptotically through spin precession frequencies. In future work, this framework can be extended to investigate how a rotating naked singularity causes the spin axis of a nearby pulsar to precess due to geodetic and frame-dragging effects. This would further allow us to explore how such spin precession influences the observed pulse frequencies on Earth, offering insight into the astrophysical signatures of naked singularities.
\par
We also found that the ISCO is absent in NNS. Orbital plane precession frequency/ nodal precession frequency in RNS shows similar behavior to the LT precession frequency caused by the frame-dragging effect, i.e., it attains a certain peak. After this, it starts decreasing as one approaches the inner edge of the disk, i.e., ISCO. This frequency vanishes before it reaches the inner edge of the disk and becomes negative, indicating a reversion in the precession direction. Similar behavior is observed in the Kerr naked singularity \cite{Chakraborty:BHandNS}. We found the best-fit values of RNS parameters for five XRBs by running MCMC simulation. Contour plots of posterior distribution also highlight the correlation between parameters, indicating a negative correlation between $M/M_\odot$ and $r/M$ parameters for all XRBs discussed. For the best estimate of the spin parameter, the $r_{\text{ISCO}}$ is closer to the central object for GRO J1655-40, XTE J1859+226, and XTE J1550-564 compared to the $r_{\text{ISCO}}$ for the Kerr spacetime as obtained in \cite{Motta:precise-mass,Motta:XTE564,Motta:XTE226}. This shows that the inner edge of the accretion disk in RNS extends much closer to the central object than the Kerr black hole.
\par
Continuum-fitting method and the analysis of the profile of the broad K$\alpha$ iron line are two methods developed to estimate the spin parameter and check the Kerr background. However, there is a disagreement between estimates of spin parameters within RPM \cite{Bambi:testing-rpm,Motta:precise-mass} and other methods for GRO J1655-40 \cite{Shafee:2005ef,gro-reflection}. It has been argued in \cite{Bambi:testing-rpm} that even considering various deformation parameters, the tension remains, suggesting that using a different black hole model is unlikely to fix the issue. In light of this, we investigated whether the tension between different spin measurement methods for GRO J1655-40 can be resolved with a naked singularity model instead of a Kerr black hole. By comparing the parameters in Table.~\ref{tab:2} and Table.~\ref{Tab.4}, our analysis indicates that even with this alternative model, the tension between the different spin measurements persists, suggesting that the possibility of resolving the disagreement through a NS model is unlikely. The difference in spin parameter estimate also persists for XTE J1859+226 and GRS 1915+105 in comparison with spectral analysis outcomes given in \cite{GRSspectralAnalys,XTE226spinSpectro}. QPOs are thought to originate from the innermost accretion flow \cite{originate}. However, in Table.~\ref{tab:2}, the emission radius for QPOs (5.48) is significantly larger than the ISCO radius (0.1069) for GRO J1655-40. This trend is consistent across all X-ray binaries (XRBs), where the emission radius is typically much greater than the ISCO radius. The observed QPOs occurring far away from the ISCO, rather than close to it, challenge the assumption that the XRBs studied in Table.~\ref{tab:2} are well-described by the RNS model. Hence, the RNS model could be ruled out for a realistic scenario. Instead, this finding may favor alternative models that better explain the observed dynamics and align spin parameter estimates with those from continuum fitting and spectral methods.
\par
It should be emphasized that so far, there has been no direct evidence about the detection of an event horizon of an astrophysical object confirming the existence of a black hole. Thus, any black hole candidate could also be a potential naked singularity mimicking the gravitational behavior of a black hole. In this connection, the shadow cast by a black hole and a naked singularity can be highly similar \cite{Kaur:2021wgy}, particularly when two different gravitational objects have the same shadow size. In certain cases, simulations suggest that naked singularities form an arc-shaped shadow which differs from a closed curve boundary of the shadow formed by black holes \cite{patel_light_2022}. However, the null- and timelike- geodesics around two objects could be different. In this case, the intensity distribution of light emitted by the accreting matter on different gravitational compact objects would be different \cite{Saurabh:2023otl}. 
Furthermore, not all naked singularities can be equal as far as observational tests are concerned. Researchers have found JMN1 naked singularity in concordance with the EHT observations of Sgr A*, whereas Reissner-Nordstr\"{o}m naked singularity is completely ruled out \cite{Vagnozzi:2022moj}. From this discussion and our results, it follows that the naked singularities should be tested alternatively with precession frequencies of different observers in strong gravitational regimes such as QPOs.
\section*{Appendix A: Derivation of General Spin Precession}
\label{appB}
\def\theequation{A.\arabic{equation}}
\setcounter{equation}{0}
In this appendix, we provide the detailed derivation of general spin precession, following the methodology outlined in \cite{Chakraborty:2016mhx}. The one form of the general spin precession frequency of a test stationary gyroscope is given by \cite{Straumann,AzregAnou2019GyroscopePF}
\begin{align}
\label{oneForm}
\tilde{\Omega}_{p}=\frac{1}{2K^2}\star(\tilde{K} {\wedge} d \tilde{K}),
\end{align}
where $\star$ is the Hodge star operator, $\wedge$ is the wedge product, and $\tilde{K}$ is the corresponding co-vector of $K$.
The general form of a stationary and axisymmetric spacetime is given by
\begin{align}
\label{genmetric}
ds^2=g_{tt}dt^2+g_{rr}dr^2+g_{\theta\theta}d\theta^2+g_{\phi\phi}d\phi^2+2g_{t\phi}dt d\phi,
\end{align}
which admits the time translation and azimuthal Killing vectors given by $\partial_{t}$ and $\partial_{\phi}$ respectively, as the metric is invariant with respect to them. The Killing vector associated with time translations yields a conserved energy for a particle moving in this spacetime, and the Killing vector associated with rotations yields a conserved angular momentum. The covector corresponding to the Killing vector $K=\partial_{t}+\Omega\partial_{\phi}$ is given by
\begin{align*}
\tilde{K}&=g_{t\mu}dx^{\mu}+\Omega g_{\gamma\phi}dx^{\gamma}.
\end{align*}
Expanding $\tilde{K}$ using indices, we get
\begin{align*}
\tilde{K}&=\left(g_{tt}dt+g_{t\phi}d\phi+g_{tb}dx^{b}\right)+\Omega \left(g_{t\phi}dt+g_{\phi\phi}d\phi+g_{b\phi}dx^{b}\right).
\end{align*}
In general, $g_{tb}=g_{\phi b}=0$ for $b=r, \theta$ for stationary axisymmetric spacetime which leads us to
\begin{align*}
\tilde{K}&=\left(g_{tt}dt+g_{t\phi}d\phi\right)+\Omega \left(g_{t\phi}dt+g_{\phi\phi}d\phi\right).
\end{align*}
Applying the exterior derivative to $\tilde{K}$ and then taking wedge product with $\tilde{K}$ we get
\begin{align*}
d\tilde{K}&=\left(g_{tt,b}dx^b {\wedge}dt+g_{\phi\phi,b}dx^b {\wedge}d\phi\right)+\Omega\left(g_{t\phi,b}dx^b {\wedge}dt+g_{\phi\phi,b}dx^b {\wedge}d\phi\right),\\
\tilde{K} {\wedge} d \Bar{K}&=\left((g_{tt}g_{t\phi,b}-g_{t\phi}g_{tt,b}+\Omega(g_{tt}g_{\phi\phi,b}-g_{\phi\phi}g_{tt,b})+\Omega^2(g_{t\phi}g_{\phi\phi,b}-g_{\phi\phi}g_{t\phi,b})\right (dt{\wedge}dx^b{\wedge}d\phi).
\end{align*}
Applying the Hodge star operator to the above expression leads us to
\begin{align}
\label{b2}
*( \tilde{K} {\wedge} d \tilde{K})=&\left[(g_{tt}g_{t\phi,b}-g_{t\phi}g_{tt,b}+\Omega(g_{tt}g_{\phi\phi,b}-g_{\phi\phi}g_{tt,b})+\Omega^2(g_{t\phi}g_{\phi\phi,b}-g_{\phi\phi}g_{t\phi,b})\right]*(dt{\wedge}dx^b{\wedge}d\phi),
\end{align}
where
\begin{align*}
*(dt{\wedge}dx^b{\wedge}d\phi)=\eta^{tb\phi d}g_{d\nu}dx^{\nu}=\frac{1}{\sqrt{-g}}\epsilon_{\phi b d} g_{d \nu}dx^{\nu}.
\end{align*}
We get $K^2$ as
\begin{align}
\label{b3}
K^2=g_{tt}+2\Omega g_{t\phi}+\Omega^2 g_{\phi\phi}.
\end{align}
After using (\ref{b2}) and (\ref{b3}), the one-form of general spin precession (\ref{oneForm}) becomes
{\small
\begin{align}
	\tilde{\Omega}_{p}=&\dfrac{\epsilon_{\phi b d} g_{b\nu}dx^{\nu} }{2 \sqrt{-g}\left(1+2\Omega \frac{g_{t\phi}}{g_{tt}}+\Omega^2\frac{ g_{\phi\phi}}{g_{tt}}\right)}\left[\left(g_{t\phi,b}-\frac{g_{t\phi}}{g_{tt}}g_{tt,b}\right)+\Omega\left(g_{\phi\phi,b}-\frac{g_{\phi\phi}}{g_{tt}}g_{tt,b}\right)\right.\left.+\Omega^2\left(\frac{g_{t\phi}}{g_{tt}}g_{\phi\phi,b}-\frac{g_{\phi\phi}}{g_{tt}}g_{t\phi,b}\right)\right].
\end{align}}%
The corresponding vector of the covector of general spin precession of a gyroscope reduces to \cite{Chakraborty:2016mhx}
{\small
\begin{align}
	\label{1.4.27}
	\Omega_\text{p}=&\dfrac{\epsilon_{\phi b d} }{2 \sqrt{-g}\left(1+2\Omega\frac{ g_{t\phi}}{g_{tt}}+\Omega^2\frac{ g_{\phi\phi}}{g_{tt}}\right)}\left[\left(g_{t\phi,b}-\frac{g_{t\phi}}{g_{tt}}g_{tt,b}\right)+\Omega\left(g_{\phi\phi,b}-\frac{g_{\phi\phi}}{g_{tt}}g_{tt,b}\right)\right.\left.+\Omega^2 \left(\frac{g_{t\phi}}{g_{tt}}g_{\phi\phi,b}-\frac{g_{\phi\phi}}{g_{tt}}g_{t\phi,b}\right)\right]\partial_{d},
\end{align}}%
where $ d, b = r, \theta$. In a strong gravity domain, we can numerically predict the spin precession frequency by transforming Eq.~(\ref{1.4.27}) from the coordinate basis to the orthonormal Copernican basis. We normalize the coordinate basis vectors as $\hat{r} = \dfrac{1}{\sqrt{g_{rr}}} \partial_r$ and $\hat{\theta} = \dfrac{1}{\sqrt{g_{\theta\theta}}} \partial_\theta$. This yields the spin precession vector in terms of $\hat{r}$ and $\hat{\theta}$, as given by (\ref{den}).

\section*{\texorpdfstring{Appendix B: Evaluating the limits $\lim_{(r\to 0^+,\,u\to 0)} \Omega_{\pm}$}{Appendix: Evaluating the limits Omega}}\label{appA}
\def\theequation{B.\arabic{equation}}
\setcounter{equation}{0}
In order to evaluate the limits $\lim_{(r\to 0^+,\,u\to 0)} \Omega_{\pm}$, we place ourselves in the $ur$-plane (axis $u$ horizontal and axis $r$ vertical). Let $r=lg(u)$, with $g(0)=0$ and $l>0$ is a parameter having the dimension of length, be a general path in the $ur$-plane that passes through the singularity at ($r=0,\,u=0$). Substituting $r=lg(u)$ into~(\ref{3.3}), we obtain
\begin{equation}\label{A1}	
\Omega_{\pm}= \frac{a M l^2g^2 (M +2 lg) \sqrt{1-u^2}\pm (lg+M) (l^2g^2+a^2 u^2) \sqrt{l^4g^4+a^2 (M+lg)^2}}{\sqrt{1-u^2} \{(M+lg)^2 (l^4g^4
	+a^4u^2)+a^2 l^2g^2 [2 M^2+4 M l g+l^2g^2 (1+u^2)]\}}.
\end{equation}

The limits $\lim_{(r\to 0^+,\,u\to 0)} \Omega_{\pm}$ exist if they are independent of the shape of the function $g(u)$ and independent of the value of the parameter $l$. Since $g(0)=0$, the function $g(u)$ behaves as $u^n$ ($n>0$) as $u\to 0$. There are three cases to envisage: (a) $n\geq 1$, (b) $\dfrac{1}{2}\leq n< 1$, and (c) $0<n<\dfrac{1}{2}$. We only consider the case (a) and the treatment of the two other cases is similar. 

If (a) $n\geq 1$, we have either $\lim_{u\to 0}g^2/u^2=1$ or $\lim_{u\to 0}g^2/u^2=0$ and $\lim_{u\to 0}g^4/u^2=0$. Examples of such paths are straight lines through the singularity $r=g(u)=lu$, or paths of the form $r=l(u + u^2)$, $r=l[u + u^2/(|u|+5)]$, $r=l(u^2 + u^4)$, $r=lu^{3/2}$, etc. In this case, we can factor $u^2$ in the numerator and denominator of~(\ref{A1}) and eliminate it to obtain
\begin{equation}\label{A2}	
\Omega_{\pm}= \frac{a M l^2\dfrac{g^2}{y^2} (M +2 lg) \sqrt{1-u^2}\pm (lg+M) \Big(l^2\dfrac{g^2}{y^2}+a^2\Big) \sqrt{l^4g^4+a^2 (M+lg)^2}}{\sqrt{1-u^2}\,\Big[(M+lg)^2 \Big(l^4\dfrac{g^4}{y^2}
	+a^4\Big)+a^2 l^2\dfrac{g^2}{y^2} [2 M^2+4 M lg+l^2g^2 (1+u^2)]\Big]}.
\end{equation}
Taking the limit $u\to 0$ and using $g(0)=0$, we obtain
\begin{equation}\label{A3}
\lim_{u\to 0}\Omega_{+}=\frac{1}{a}\,,
\end{equation}
\begin{equation}\label{A4}
\lim_{u\to 0}\Omega_{-}=
\begin{cases}
	-\dfrac{1}{a}\,, & \text{if }\lim_{u\to 0}\dfrac{g^2}{u^2}=0\vspace{3mm}\\
	-\dfrac{a}{a^2+2l^2}\,, & \text{if }\lim_{u\to 0}\dfrac{g^2}{u^2}=1
\end{cases}\,.
\end{equation}
This shows clearly that $\lim_{(r\to 0^+,\,u\to 0)} \Omega_{-}$ does not exist, since, for instance, on straight lines $r=lg(u)=lu$, where $n=1$, the limit $-a/(a^2+2l^2)$ depends on the slope $l$ of the line. Moreover, on paths with $n>1$, such as the path $r=lu^{3/2}$, the limit is $-1/a$ and is different from that on straight lines having a non-vanishing slope [which is $-a/(a^2+2l^2)$].

For the other cases, (b) $\dfrac{1}{2}\leq n< 1$ and (c) $0<n<\dfrac{1}{2}$, we obtain
\begin{equation*}
\lim_{u\to 0}\Omega_{+}=\frac{1}{a}\,,
\end{equation*}
\begin{equation*}
\lim_{u\to 0}\Omega_{-}=0\,,
\end{equation*}
which shows again that $\lim_{(r\to 0^+,\,u\to 0)} \Omega_{-}$ does not exist and that in all three cases, we have $\lim_{(r\to 0^+,\,u\to 0)} \Omega_{+}=1/a$.
\section*{Appendix C: Weak field limit of Rotating Naked Singularity}
\label{appC}
\def\theequation{C.\arabic{equation}}
\setcounter{equation}{0}
In this appendix, we derive the weak-field limits of the Kerr and RNS metrics and verify their agreement in the asymptotic regime \( r \gg M \). \newline
Starting with the Kerr metric ~\eqref{4.1}, we expand the metric components to leading order in \( \frac{M}{r} \) and \( \frac{a}{r} \), assuming \( r \gg M \). The resulting approximations are
\begin{align}
	g_{tt} &\approx -\left(1 - \frac{2M}{r} \right), &
	g_{t\phi} &\approx -\frac{4Ma \sin^2\theta}{r}, \notag \\
	g_{rr} &\approx 1 + \frac{2M}{r}, &
	g_{\theta\theta} &\approx r^2, \notag \\
	g_{\phi\phi} &\approx r^2 \sin^2\theta. \label{C.1} 
\end{align}
For RNS metric as given in Eq.~\eqref{2.1}, using the same weak-field assumption \( r \gg M \), the metric functions simplify to
\begin{align}
	f \approx Mr, \quad \rho^2 \approx r^2, \quad \Delta \approx \frac{r^2} {\left(1 + \frac{2M}{r} \right)}, \quad \Sigma \approx r^4. \label{C.2}
\end{align}
Substituting into the metric and expanding to leading order gives
\begin{align}
	g_{tt} &\approx -\left(1 - \frac{2M}{r} \right), &
	g_{t\phi} &\approx -\frac{4aM \sin^2\theta}{r}, \notag \\
	g_{rr} &\approx 1 + \frac{2M}{r}, &
	g_{\theta\theta} &\approx r^2, \notag \\
	g_{\phi\phi} &\approx r^2 \sin^2\theta. \label{C.3}
\end{align}
By comparing the metric components in ~\eqref{C.1} and \eqref{C.3}, we observe that both the Kerr and RNS metrics reduce to the same weak-field form as
\begin{align}
	ds^2 \approx -\left(1 - \frac{2M}{r} \right) dt^2 - \frac{4aM \sin^2\theta}{r} dt\, d\phi + \left(1 + \frac{2M}{r} \right) dr^2 + r^2 d\theta^2 + r^2 \sin^2\theta\, d\phi^2. \label{C.4}
\end{align}
This confirms that the RNS metric is consistent with the Kerr metric in the limit \( r \gg M \).

\subsection*{Acknowledgments}
We would like to thank Chandrachur Chakraborty for his insightful comments during the preparation of this manuscript. MJ would like to thank Qiang Wu and Tao Zhu for several fruitful discussions related to this paper. He would like to thank Zhejiang University of Technology, Hangzhou, for providing hospitality where part of this work was completed.

\end{document}